\documentclass[a4paper]{aa} 

\usepackage{graphicx}
\usepackage{txfonts}

\usepackage{caption}
\usepackage{amsmath}
\usepackage{hyperref}
\usepackage[utf8]{inputenc}

\usepackage{indentfirst}

\usepackage{lineno}

\usepackage{amsmath}
\usepackage{lscape}

\usepackage{xcolor}

\newcommand{\violet}[1]{\textcolor{violet}{#1}}

\begin{document} 

\title{Vorticity and magnetic dynamo from subsonic expansion waves \\ II. Dependence on the magnetic Prandtl number, forcing scale, and cooling time}

\authorrunning{Elias-López et al.}
\titlerunning{Vorticity and magnetic dynamo from subsonic expansion waves}
   \author{Albert Elias-López
          \inst{1,2},
          Fabio Del Sordo \inst{1,2,3}
          and
          Daniele Viganò \inst{1,2,4}
          }

\institute { Institute of Space Sciences (ICE-CSIC), Campus UAB, Carrer de Can Magrans s/n, 08193, Barcelona, Spain
\and
Institut d’Estudis Espacials de Catalunya (IEEC), 08034 Barcelona, Spain
\and
INAF, Osservatorio Astrofisico di Catania, via Santa Sofia, 78 Catania, Italy
\and
Institute of Applied Computing \& Community Code (IAC3), University of the Balearic Islands, Palma, 07122, Spain
\\
\email{albert.elias@csic.es}\\
\email{delsordo@ice.csic.es}\\
\email{daniele.vigano@csic.es}
             }
   \date{Received ---------; accepted --------}

    \abstract
  % context heading (optional)
   {The amplification of astrophysical magnetic fields takes place via dynamo instability in turbulent environments. Vorticity is usually present in any dynamo, but its role is not yet fully understood.
   }
  % aims heading (mandatory)
   {This work is an extension of previous research on the effect of an irrotational subsonic forcing on a magnetized medium in the presence of rotation or a differential velocity profile. We aim to explore a wider parameter space in terms of Reynolds numbers, the magnetic Prandtl number, the forcing scale, and the cooling timescale in a Newtonian cooling. We studied the effect of imposing that either the acceleration or the velocity forcing function be curl-free and evaluated the terms responsible for the evolution vorticity.}
  % methods heading (mandatory)
   {We used direct numerical simulations to solve the fully compressible, resistive magnetohydrodynamic equations with the Pencil Code. We studied both isothermal and non-isothermal regimes and addressed the relative importance of different vorticity source terms.}
  % results heading (mandatory)
   {We report no small-scale dynamo for the models that do not include shear. We find a hydro instability, followed by a magnetic one, when a shearing velocity profile is applied. The vorticity production is found to be numerical in the purely irrotational acceleration case. Non-isothermality, rotation, shear, and density-dependent forcing, when included, contribute to increasing the vorticity.}
   % conclusions heading
   {As in our previous study, we find that turbulence driven by subsonic expansion waves can amplify the vorticity and magnetic field only in the presence of a background shearing profile. The presence of a cooling function makes the instability occur on a shorter timescale. We estimate critical Reynolds and magnetic Reynolds numbers of 40 and 20, respectively.}


\maketitle

\section{Introduction} 

\noindent The occurrence of vortical flows is of high importance in astrophysics due to their close connection with turbulence. Vorticity is a way to characterize turbulent flows, which are ubiquitous in astrophysical contexts. However, the connection between turbulence and the processes responsible for the amplification of astrophysical magnetic fields has not been fully elucidated. Magnetic fields are amplified at all scales in various astrophysical environments, from planets to the extragalactic medium. While the range of scales, densities, and velocities is fairly wide, the instability responsible for the amplification of magnetic fields, known as dynamo, is thought to be active in all of these environments. Numerical models are commonly used to simulate many of these astrophysical contexts, from planets \citep[e.g.,][]{Jones2011} to galaxies \citep[see, e.g., ][for a recent review]{Brandenburg&Ntormousi2023}, and in each of them the turbulence is injected through a forcing mechanism. In this study we focus on two open questions, namely (a) whether the occurrence of dynamo instability depends on the forcing mechanism, and (b) the minimum ingredients needed to trigger a dynamo in a magnetohydrodynamic (MHD) turbulent medium.

In our attempt to address these two points, we expand on the work by \cite{MeeBrandenburg2006}, \cite{DelSordo&Brandenburg2011}, and \cite{EliasLopezetal2023}, who made use of a purely irrotational (i.e., curl-free) forcing of the velocity field to create a turbulent medium. The used forcing function mimics the occurrence of spherical expansion waves in an initially homogeneous medium, in a subsonic regime. The aforementioned studies found no vorticity amplification from the forcing alone in an isothermal purely hydrodynamic (HD) environment, and no dynamo amplification unless a background shearing profile is added to the system, independently of the equation of state. A similar result was obtained by \cite{Kahniashvilietal2012}, who employed a curl-free forcing function to model inflationary scenarios in the early Universe. They found that the Lorentz force may produce some vorticity in isothermal models, but the magnetic field eventually dissipates without undergoing any kind of amplification. \cite{Dosopoulouetal2012} found analogous results in the context of magnetized and rotating cosmological models. Nevertheless, other approaches found vorticity and magnetic fields to be exponentially amplified when a purely irrotational forcing is added in the form of a stochastic function in Fourier space \citep[e.g.,][]{Federrathetal2010,AchikanathFederrath2021,Seta&Federrath2022}, even in isothermal contexts, in the absence of large-scale contributions to the forcing, such as rotation or shear. As a consequence, it remains unclear which are the minimum ingredients needed to excite a dynamo. It is possible that the difference between these different studies resides either in the used forcing function or in the exploration of different regions of a fairly wide parameter space.\ This open question in particular motivated the work presented here.

In the present study we took as a starting point the work done by \cite{EliasLopezetal2023}, and we investigated the effect of varying the explosion width and the magnetic Prandtl number, as well as the effect of using a Newtonian cooling function, on the amplification of the vorticity and magnetic field. We also studied the difference between (a) forcing the turbulence by imposing an exactly irrotational forcing for the acceleration and (b) a variant that forces a locally fully potential velocity field and hence may include vorticity created by density fluctuations. The article is organized as follows. In Sect. \ref{sec:model} we present the numerical model used to perform the study, in Sect. \ref{sec:results} we describe the main results, and in Sect. \ref{sec:conclusions} we discuss some of the conclusions of our work.

\section{Model and numerical methods}\label{sec:model}
\subsection{MHD equations with rotation and shear}
\label{sec:mhdeq}

\noindent We employed the same numerical models as in \cite{EliasLopezetal2023}, which for completion we briefly explain below. We used the public 3D MHD code Pencil Code\footnote{https://github.com/pencil-code} \citep{ThePencilCode}, which is a non-conservative, high-order, finite-difference code (sixth-order accurate in space and third-order Runge-Kutta in time). We solved the nonideal fully compressible MHD equations following an approach similar to what was done by \cite{DelSordo&Brandenburg2011}, that is, either in a rigidly rotating frame, with angular velocity $\boldsymbol{\omega}=\Omega \mathbf{e}_z$, or with a differential velocity (shear) given by $\mathbf{u}^S = u_y^S(z) \mathbf{e}_y$, with $u_y^S = A \cos (k z)$, similar to \cite{Skoutnev2022MNRAS}, \cite{Kayplaetal2009sinusoidal}, and \cite{Kayplaetal2009}. In our models, z ranges from $-\pi$ to $\pi$, so when we included a shearing profile we set $k = 1$. This allows simple periodic boundary conditions in the three directions.  While in \cite{EliasLopezetal2023} we explored the role played by the shear amplitude, in the present work we used $A=0.2$ in all the shearing cases. 

The set of equations employed consists of the continuity equation, the momentum equation, the entropy equation (which we only solve for the non-isothermal/baroclinic case), and the induction equation, respectively: 
\begin{equation}
  \dfrac{D \ln \rho}{D t} = - \nabla \cdot \mathbf{u}~, \label{Continuity_equation}
\end{equation}
\begin{equation}
 \dfrac{D \mathbf{u}}{D t} = -\frac{\nabla p}{\rho} + \mathbf{F}_{visc} + \frac{\mathbf{J} \times \mathbf{B}}{\rho} - 2 \boldsymbol{\Omega} \times \mathbf{u} + \mathbf{f} + \mathbf{f}_s, \label{Equation_of_motion} 
 \end{equation}
\begin{equation}
 T \dfrac{D s}{D t} = 2 \nu \boldsymbol{S} \otimes \boldsymbol{S} + \rho^{-1} \nabla (c_p \rho \chi \nabla T) + \rho^{-1} \eta \mu_0 \boldsymbol{J}^2 - \frac{1}{\tau_{cool}}(c_s^2 - c_{s0}^2)~, \label{Entropy_equation_Pencil}
 \end{equation}
\begin{equation}
 \dfrac{\partial \mathbf{A}}{\partial t} = \mathbf{u}\times (\nabla \times \mathbf{A}) + \eta \nabla^2 \mathbf{A}~.
\label{Induction equation}
\end{equation}
In Eqs. \ref{Continuity_equation}-\ref{Induction equation} $\rho$ is the mass density; $\mathbf{u}(t)=(\mathbf{u}^S + \mathbf{u}'(t))$ is the total velocity, which can be thought of as the sum of shearing and turbulent velocities; $p$ is the pressure; $\mathbf{B}$ the magnetic field; $\mathbf{A}$ its vector potential (i.e., $\mathbf{B} = \nabla\times \mathbf{A}$); $\mathbf{J}=(\nabla \times \mathbf{B})/\mu_0$ is the electrical current density (where $\mu_0$ is the vacuum permeability); $\mathbf{F}_{visc} = \rho^{-1} \nabla \cdot (2\rho \nu \mathbf{S})$, where the traceless rate of strain tensor $\mathbf{S}$ has components $S_{ij} = (1/2)(u_{i,j}+u_{j,i} - (1/3) \delta_{ij} \nabla \cdot \mathbf{u})$; $\mathbf{f}$ the expansion wave forcing (see below for their definitions); $\chi$ is the thermal diffusivity; $\eta$ is the magnetic diffusivity, $c_{s0}$ is the initial, uniform sound speed (proportional to the initial temperature) and $\tau_{cool}$ is the cooling term timescale, introduced to avoid an indefinite heating. The advective derivative operator is $D/Dt := \partial / \partial t + \mathbf{u} \cdot \nabla$. The differential velocity profile is imposed directly on the y component of the total velocity by an amount proportional to the difference of velocity and the profile itself, so  $u_y$ does not deviate much from the differential shearing profile:
\begin{equation*}
    \mathbf{f}_s = \frac{1}{\tau_S} (u_y^S - u_y) \mathbf{\hat{y}} ~,
\end{equation*}
where we fix $\tau_S=1$ (as for $\tau_{cool}$ for the entropy, the smaller the value, the more effective is the keeping $u_y$ close to $u_y^S$).

In order to close the system of equations, we consider two types of equation of state (EoS): (1) a simple barotropic EoS $p(\rho)=c_s^2 \rho$, where we fix the value of the sound speed $c_s = 1$, or (2) an ideal EoS, also dubbed the baroclinic case, $p(\rho,T)=\rho R_g T$, with $R_g$ the specific gas constant and $T$ the temperature; in this case, the sound speed squared is $c_s^2 = (\gamma - 1)c_p T$, where we fix the adiabatic index $\gamma = c_p/c_v = 5/3$ (corresponding to a monatomic perfect gas), and $c_p$ and $c_v$ are the specific heats at constant pressure and constant volume, respectively.

The forcing is applied as an acceleration in the momentum equation and we explore two different ways: either the acceleration itself is irrotational, as in \cite{EliasLopezetal2023}, or the force itself is irrotational:
\begin{align}
  \mathbf{f_{\text{acc}}}(\mathbf{x},t) = \nabla \phi(\mathbf{x},t) = K \nabla e^{ -(\mathbf{x} - \mathbf{x_f}(t))^2 /R^2 }, \quad \text{or} \label{Forcing} \\
  \mathbf{f_{\text{mom}}}(\mathbf{x},t) =  \frac{\rho_0}{\rho} \nabla \phi(\mathbf{x},t) = K\frac{\rho_0}{\rho} \nabla e^{ -(\mathbf{x} - \mathbf{x_f}(t))^2 /R^2 } \quad \quad, \text{  } \label{Forcing momentum}
\end{align}
where $\mathbf{x}_f(t)$ is the randomly changed expansion wave center, $\rho_0$ is the  initial, uniform density, $K =\phi_0 \sqrt{R /(c_{s0} \Delta t)}$ is the normalization factor, $c_{s0}$ is the initial sound speed, $R$ is the radius of the Gaussian, $\Delta t$ is the time interval after which a new expansion wave is forced in a new position (and it can be as short as the time step), and $\phi_0$ controls the overall forcing amplitude and has dimensions of velocity squared. In either case, the associated forcing  representative wavenumber is thus $k_f = 2/R$. A spherical symmetric explosion in the interstellar medium (ISM) is arguably better modeled by the second choice, since the expansion takes ISM density  fluctuations into account. We note that, in any case, this is a simplification, since important deviation from spherical symmetry are expected in detailed 3D supernova (SN)   explosion simulations \citep{Reichertetal2023}. However, in this work we aim at exploring the difference between the two kinds of forcing, within the spherical symmetric assumption, to understand if the density fluctuations alone can be responsible of vorticity-triggered dynamo action.

The simulation domain consists of a uniform, cubic grid mesh $[-\pi,\pi]^3$, with triply periodic boundary conditions. We also studied resolution convergence varying from 32$^3$ up to 256$^3$ mesh points, and we find that 256$^3$ mesh points are enough to assess our problem. We ran some models at an even higher resolution, 512$^3$, to double-check the validity of the results obtained at lower resolutions. However, simulations at this resolution are computationally expensive, so we limited their use to double check numerical convergence. The chosen velocity profile allows simple periodic boundary conditions in the z direction if $k$ is an integer. We adopted nondimensional variables by measuring speed in units of the initial sound speed, $c_{s0}$, and length in units of $1/k_1$, where $k_1$ is the smallest wave number in the periodic domain, implying that the nondimensional size of the domain is $(2\pi)^3$.

As the initial conditions: pressure and  density are set constant and with value of 1 throughout the box (making $\rho_0=1$ for all times), and so are entropy and temperature in the baroclinic case; the fluid is initially at rest so the flow is described by $\mathbf{u}=0$; the initial magnetic field is a weak seed randomly generated with an amplitude of 10$^{-6}$ in code units, uncorrelated at each point for the three components, corresponding to a $E_k \sim k^4$ power law, as reported by \cite{MeeBrandenburg2006}.

\subsection{Diagnostics}

\noindent After an initial transitory phase, the simulations reach a stationary state, over the course of which the main average quantities maintain a saturated value. In particular, we looked at the root mean square of the total velocity, $u_{rms}$. We in turn used this to define the fundamental timescale of our problem, which we call turnover time, as 
\begin{equation}
t_{turn} = (k_f u_{rms})^{-1}~.
\end{equation}
The turnover time can be understood as the average time for the fluid to cross an explosion width. In the cases where we employed the shearing profile, we also used a similar definition with the shearing wavelength, $k,$ and amplitude, $A$:
\begin{equation}
t_{shear} = (k A)^{-1}~.
\end{equation}

The root mean square values of velocity $u_{rms}$ and vorticity $\omega_{rms}$ (see Sect. \ref{Subsection: vorticity}) are used to define the following dimensionless numbers:
\begin{eqnarray*}
&&    \text{Re} = \frac{u_{rms}}{\nu k_f}~, \quad \quad \text{Rm} = \frac{u_{rms}}{\eta k_f}~, \quad \quad \text{Re$_\omega$} = \frac{\omega_{rms}}{\nu k_f^2}~,\\
&& \text{Ma}=\frac{u_{rms}}{c_s}~, \quad \quad  \text{Pm}=\frac{\nu}{\eta}~,  \quad \quad  k_{\omega}=\frac{\omega_{rms}}{u_{rms}}~,
\end{eqnarray*}
which are the Reynolds number, magnetic Reynolds number, vorticity Reynolds number, Mach number, magnetic Prandtl number, and a measure of vortical wavelength, respectively.

\subsection{Vorticity equation}
\label{Subsection: vorticity}

To study the evolution of the vorticity ($\boldsymbol{\omega} = \nabla \times \mathbf{u}$), its sources or dissipative terms, we looked at some diagnostic quantities derived from the terms of the vorticity evolution equation:

\begin{equation}
\label{Eq: Vorticity evolution}
\begin{aligned}
  \frac{\partial \boldsymbol{\omega}}{\partial t} = \boldsymbol{\nabla} \times (\mathbf{u} \times \boldsymbol{\omega}) + \boldsymbol{\nabla} \times \mathbf{F}_{visc} + \frac{\nabla \rho \times \nabla p}{\rho^2} + \nabla \times \left(\frac{\mathbf{J} \times \mathbf{B}}{\rho}\right) \\ - 2 \nabla \times ( \boldsymbol{\Omega} \times \mathbf{u}) + \nabla  \times \mathbf{f}  + \nabla  \times \mathbf{f}_S ~. %- \nabla \times \left( u_z\frac{\partial u_y^S}{\partial z}\mathbf{\hat{y}} \right)
\end{aligned}
\end{equation}
Here the first term on the right-hand side is analogous to the advective term $\nabla\times (\mathbf{u} \times \mathbf{B})$ in the induction equation, the second term represents the viscous forces acting on the system, the third is the baroclinic term, related to the EoS, the forth is the effect of the Lorentz force, the fifth appears if the system is rotating, the sixth is due to the effect of the implemented forcing and the last one regards the sinusoidal shearing profile.

Taking the dot product with $\boldsymbol{\omega}$, integrating over the volume, and using the vector identities $(\boldsymbol{\nabla} \times \mathbf{a}) \cdot \mathbf{b} = \boldsymbol{\nabla} \cdot (\mathbf{a} \times \mathbf{b}) + \mathbf{a} \cdot(\boldsymbol{\nabla} \times \mathbf{b})$ and $\nabla^2\mathbf{a} = \boldsymbol{\nabla} ( \boldsymbol{\nabla} \cdot \mathbf{a})  - \boldsymbol{\nabla} \times (\boldsymbol{\nabla} \times \mathbf{a}),$  we obtain

\begin{equation}
\begin{aligned}
  \frac{1}{2}\frac{\partial }{\partial t} \langle \boldsymbol{\omega}^2 \rangle = \langle (\mathbf{u} \times \boldsymbol{\omega}) \cdot \mathbf{q} \rangle - \nu \langle |\mathbf{q}|^2 \rangle + 2\nu\langle \mathbf{S} \boldsymbol{\nabla} ln \rho \cdot \mathbf{q} \rangle - \\ - \langle (\boldsymbol{\nabla}T \times \boldsymbol{\nabla}s) \cdot \boldsymbol{\omega} \rangle + \left\langle \frac{\mathbf{J} \times \mathbf{B}}{\rho} \cdot \mathbf{q} \right\rangle - 2\Omega\langle (\boldsymbol{e_z} \times \mathbf{u}) \cdot \mathbf{q} \rangle + \\ +\langle \mathbf{f} \cdot \mathbf{q}\rangle + \langle \mathbf{f}_S \cdot \mathbf{q}\rangle~,
\end{aligned}
\label{Eq: Vorticity source terms}
\end{equation}
where $\mathbf{q} = \boldsymbol{\nabla} \times \boldsymbol{\omega}$. We did not evolve $\boldsymbol{\omega}$ itself; thus, all the quantities are obtained from the evolution of $\mathbf{u}$. We used these diagnostic magnitudes to discriminate which are the most relevant vorticity amplification or destruction terms. Additionally,     we can address whether the total sum of terms is similar to the numerical time derivative of $\langle \boldsymbol{\omega}^2\rangle$.

\section{Results}
\label{sec:results}
\noindent The main result of our study is that, throughout our fairly wide exploration of the parameter space in terms of forcing scales (see Sects. \ref{Section: Forcing width noshear} and \ref{Section: Forcing width shear}), magnetic Prandtl number (Sect. \ref{Section: Pm}), cooling times (Sect. \ref{Section: cooling}), and Reynolds numbers, we do not obtain an HD or MHD instability unless a background shearing flow is imposed. This result holds both with a forcing acting on the momentum, as in Eq. \ref{Forcing momentum}, and on the velocity field alone, as in Eq. \ref{Forcing}.  The models including a shearing profile instead develop an exponential increase in vorticity, followed by an exponential increase in the magnetic field, unless the scale of the forcing is too small, as explained in Sect. \ref{Section: Forcing width shear}. In Sect. \ref{Subsection: vorticity} we describe the contribution of vorticity source terms from Eq. \ref{Eq: Vorticity evolution}. In Appendix \ref{App: Tabulated runs} we show all tabulated runs and diagnostics described in this work.

\begin{figure*}[t]
\includegraphics[width=.45\textwidth]{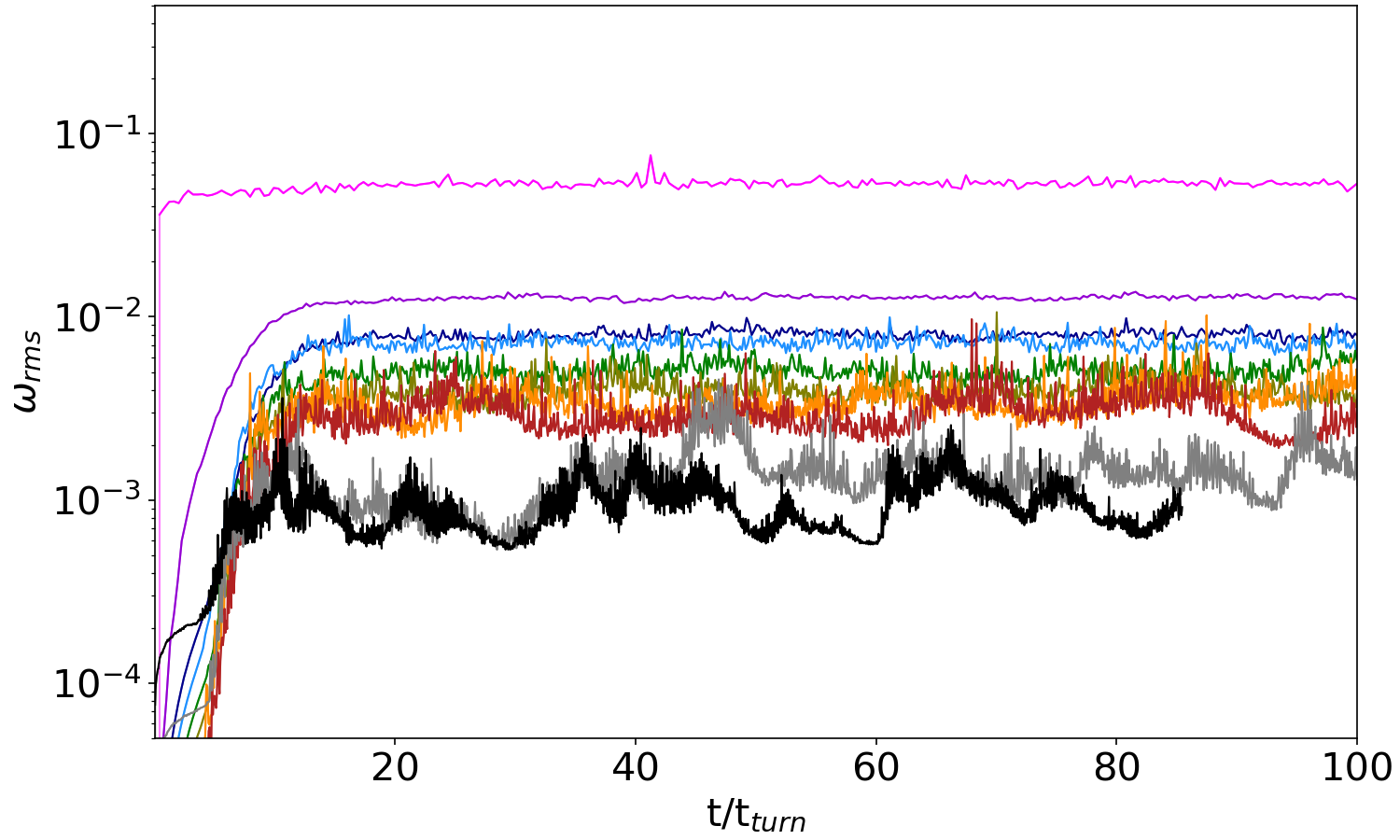}
\includegraphics[width=.5\textwidth]{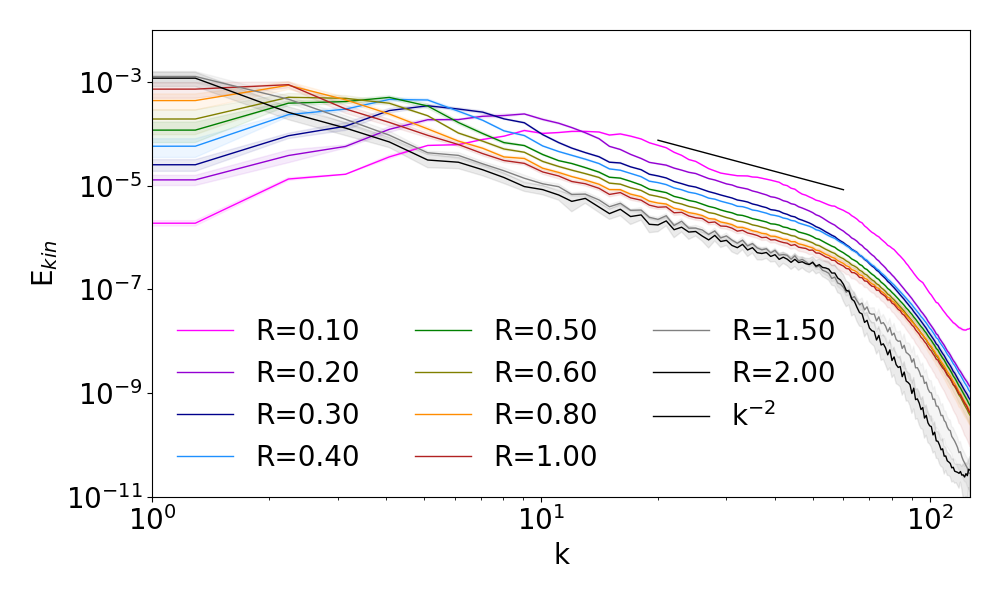}
\caption{Time evolution for $\omega_{rms}$ (left) and time-averaged kinetic spectra at saturation (right) for nonrotating isothermal runs with different explosion widths and similar total energy (see Table \ref{Tab: Barotropic runs}). Notice that the peaks in $k$ are close to the corresponding forcing wavenumber $k_f = 2/R$.}
\label{orms widths}
\end{figure*}

\subsection{Dependence on the forcing scale without shear}
\label{Section: Forcing width noshear}

\noindent We explored the role played by the forcing scale $R$,  which can be interpreted as the physical scale of the interaction between a SN explosion and the ISM. As a reference in this sense, the typical sizes of a SN remnant are typically 20-50 pc \citep{Franchettietal2012,Asvarov2014}. The size of the box can be converted into physical scales if the shear is included: for the values that we employed here, the box corresponds to about 500 pc \citep{EliasLopezetal2023}. Anyways, we stress the aim of our box simulations is to investigate basic aspects of vorticity dynamo, rather than a detailed model of the effect of SN on the ISM. 

We note that when we changed $R$, we changed $\phi_0$ accordingly in order to reach a similar $u_{rms}$. This allowed us to compare simulations with different Reynolds numbers (which scale with $R$), different energy injection scales, and different filling factors within the box. This is especially relevant when the shear is included, since it represents another physical scale and the vorticity production and dynamo depend on both (see Sect. \ref{Section: Forcing width shear}).

In the left panel of Fig. \ref{orms widths} we plot the temporal evolution of  $\omega_{rms}$ in terms of turnover time for different, representative runs without rotation, with isothermal conditions. Independently of the forcing scale $R$, vorticity reaches a steady state after less than 15 turnover times, with the exception of the smallest value of $R=0.10$, which takes less than 5 turnover times.  In the model with a forcing scale of $R=0.10,$ we see the development of local transonic flows as a consequence of the attempt to reach values of the kinetic energy similar to the cases with larger forcing widths. This is accomplished by increasing the parameter  $\phi_0$ in Eq. \ref{Forcing} and it leads to a different behavior of these models. The mean value of vorticity is observed to decrease with $R$, while its fluctuations do increase. 

In the right panel of Fig. \ref{orms widths}, we show the different kinetic spectra obtained from such runs. The corresponding forcing widths change the forcing wavelength $k_f$, thus changing the inertial range interval, as $\nu$ is kept constant. The slope of -2 seems to be independent of $R$, as already seen by \cite{MeeBrandenburg2006}. No dynamo was observed in these runs despite having reached $Rm$ larger than 200. When rotation is added, the results are the similar to \cite{EliasLopezetal2023}: we obtain steeper slopes but a similar inertial range behavior as the nonrotating cases here plotted.

\begin{figure}[h]
\centerline{\includegraphics[width=\hsize]{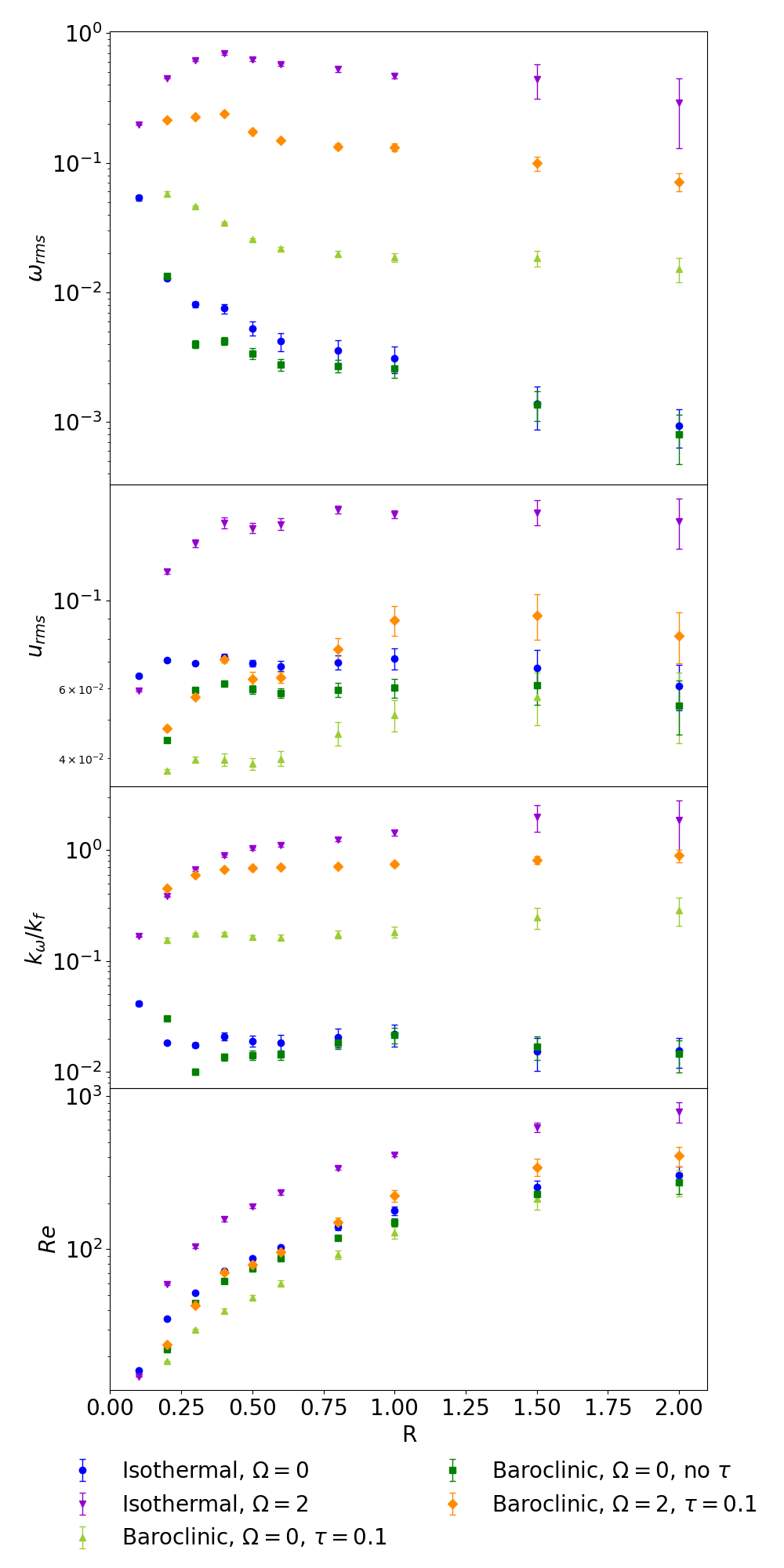}}
\caption{Different diagnostic quantities, $\omega_{rms}, u_{rms}, k_{\omega}/k_{f}$, and Re (from top to bottom), as a function of explosion width, $R$, for the forcing $f_{acc}$. These runs have Pm set to 1, i.e., Rm=Re, with a grid size of 256$^3$ and no dynamo present. The data are presented in Tables \ref{Tab: Barotropic runs} and \ref{Tab: Baroclinic runs}.}
\label{Omega rms vs width}
\end{figure}

This can be seen more clearly in Fig. \ref{Omega rms vs width}, where we plot the average of various diagnostics calculated during the saturated stage for the isothermal and baroclinic models, with or without rotation, using the forcing $f_{acc}$. We note that for $R=0.1$ the non-isothermal runs are not shown because they reach locally supersonic flows and remain numerically stable for too few time steps. Despite attempting to select a $\phi_0$ that makes $u_{rms}$ approximately independent of R, we succeeded in doing so only for the isothermal nonrotating case, while for the other non-isothermal or rotating cases $u_{rms}$ mildly depends on R. Therefore, the more important dimensionless quantity is $k_{\omega}/k_f$, as it is a measure of vorticity normalized with velocity and forcing wavelength. Figure \ref{Omega rms vs width} also illustrates many other features. First, we see how the Reynolds number increases with width for all cases, but no instability is found nevertheless. We notice that the nonrotating cases seem to have a lower overall $\omega$, but it tends to growth with smaller $R$. In contrast, for the rotating runs there is a maximum at $R=0.4$ (i.e., when expansion waves are about a fifth of the simulation domain, $k_f=5$). We also see how isothermal rotating case show the highest values for $\omega_{rms}$, $u_{rms}$, and $Re$, but in $k_{\omega}/k_f$ they are closely matched by the non-isothermal rotating with the addition of a  the thermal cooling time. As expected, the cooling term creates an additional source of dissipation, which we had to compensate for by doubling the values of $\phi_0$ (for $\tau_{cool}=0.1$). 

\begin{figure}[h]
\centerline{\includegraphics[width=\hsize]{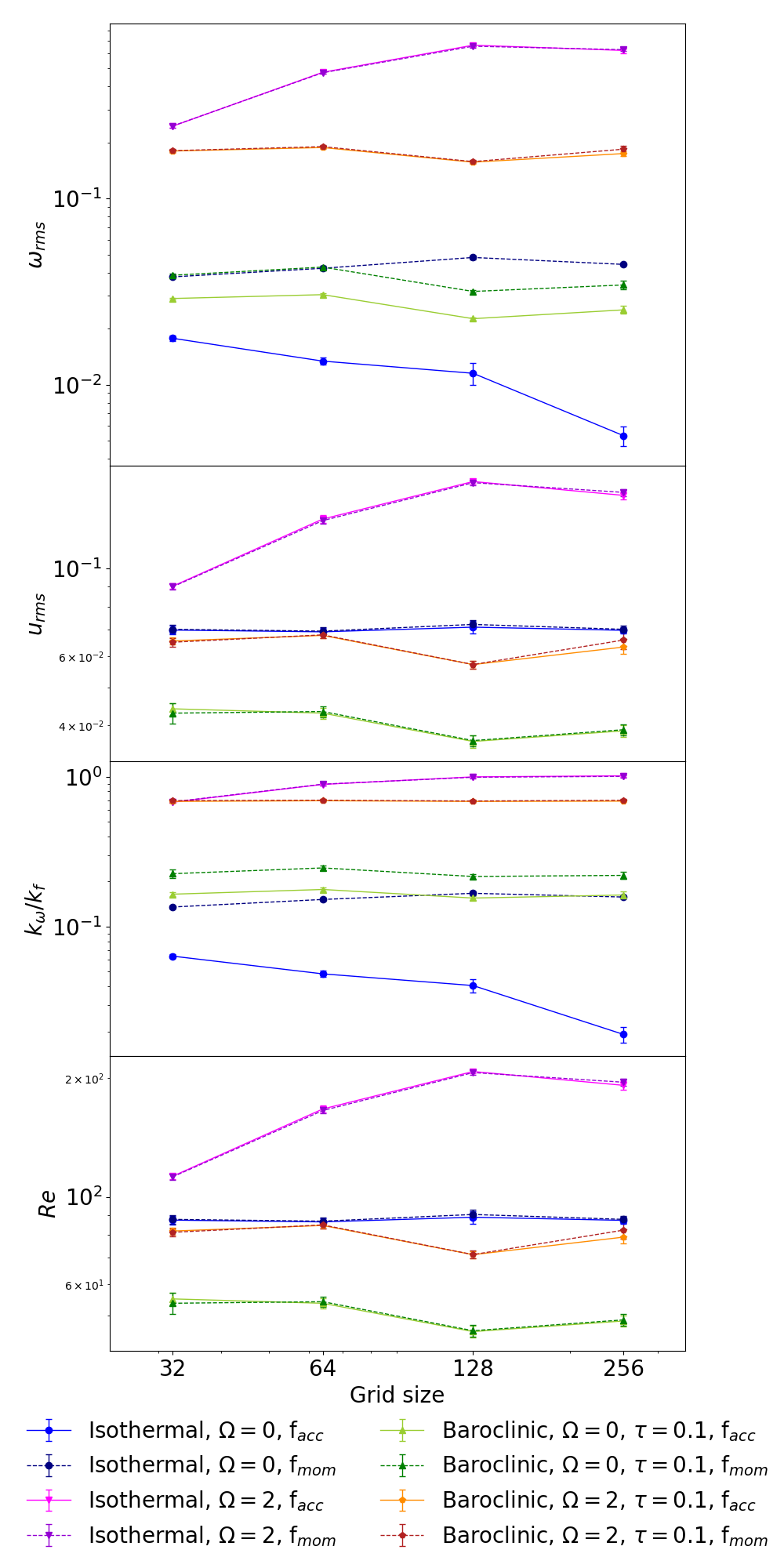}}
\caption{Same magnitudes as in Fig. \ref{Omega rms vs width} but now for runs with $R=0.5$, i.e., $k_f=4$, different resolutions, and varying the type of forcing.}
\label{Omega rms vs resolution}
\end{figure}

In Fig. \ref{Omega rms vs resolution} we plot the same quantities as in the left panel of Fig. \ref{Omega rms vs width}, comparing runs with different resolutions and keeping $R$ constant with a value of $0.5$. Most quantities seem to be resolution independent even when we move to resolution low enough to erase a good part of the inertial ranges. We observe that $u_{rms}$ is more or less resolution independent for all except two cases: (i) the isothermal nonrotating case, and (ii) the isothermal rotating case. The first case is compatible with the idea of vorticity being created solely by numerical sources.

As expected the contribution of rotation to $\omega$ is much greater than the baroclinc one in non-isothermal models, but this difference diminishes at small values of the forcing scale $R$. No dynamo is found either for the cases with rotation and/or non-isothermality with an ideal gas law and different cooling times. Thus, the results of \cite{EliasLopezetal2023} still hold when $k_f$ grows nearly up to 1, and for Rm values of several hundred.

\subsection{Dependence on the forcing scale in the presence of shear}
\label{Section: Forcing width shear}

\noindent The general behavior with the presence of shear is similar to that found in \cite{EliasLopezetal2023}: an HD instability develops first with an exponential growth of $\omega_{rms}$, which is then closely followed by a magnetic instability leading to an exponential amplification of $b_{rms}$. The complete set of runs with the instability is in Table \ref{Tab:  Shear runs} alongside with their diagnostics. After the linear phase of the dynamo a winding phenomena is seen for all cases, independently of $R$, Prandtl number and resolution. During this process $B_y$ is further amplified in the shearing direction in a linear way, by winding.

\begin{figure}[h]
\centerline{\includegraphics[width=\hsize]{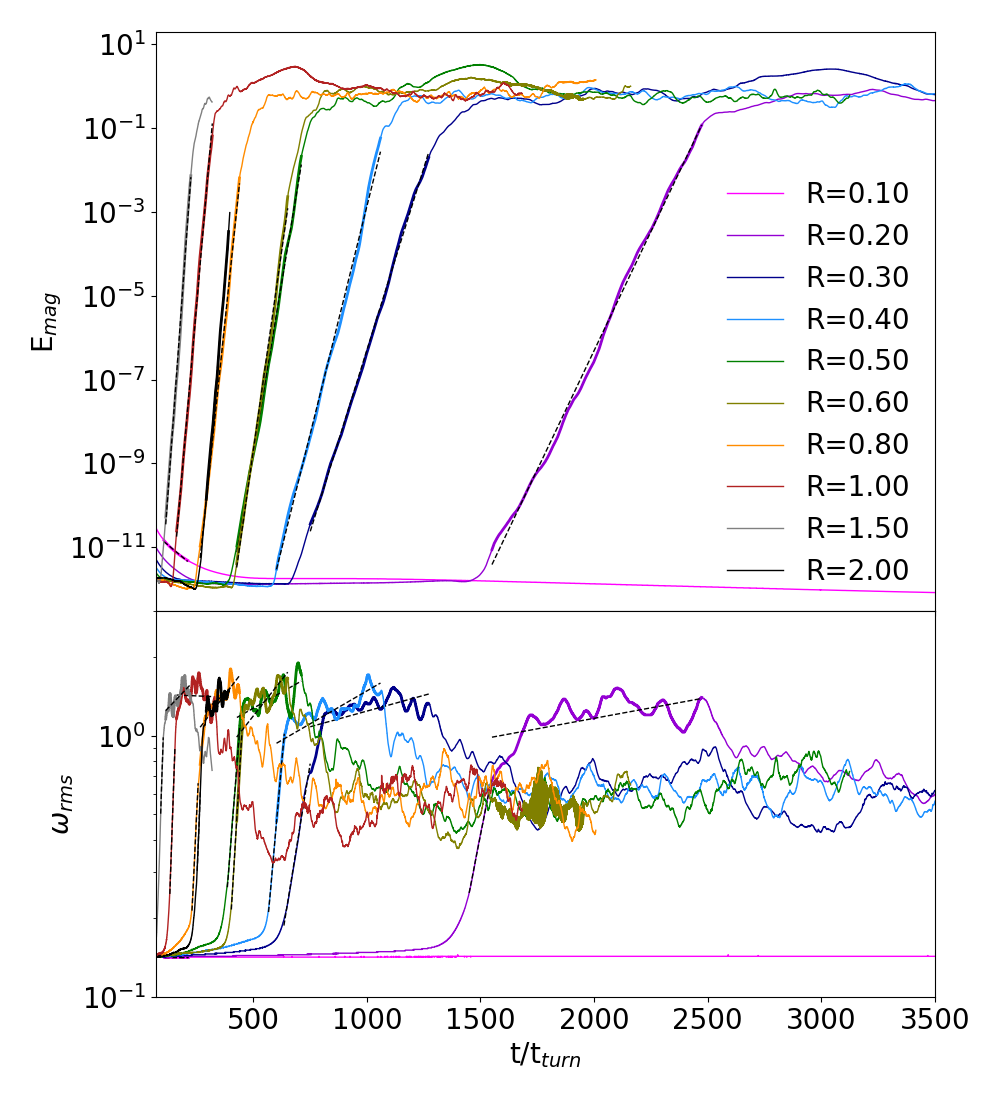}}
\caption{Time evolution (in units of $t_{turn}$) of vorticity and magnetic energy of runs with the shearing profile and different $R$ (see Table \ref{Tab:  Shear runs}). If we use the shearing timescales, the magnetic instability is between 70 and 200 $t_{shear}$ with more similar growth rates (see Fig. \ref{Fig: Growth rates vs R}). In both cases, $R=1.00$ and $R=1.50$ take the least amount of time to reach the instability. Dashed lines represent the exponential fits for $E_{mag}$ during dynamo growth (top panel), $\omega_{rms}$ during vorticity growth, and $\omega_{rms}$ during dynamo growth (bottom panel).}
\label{Fig: Timeseries growth rates}
\end{figure}

\begin{figure}[h]
\centerline{\includegraphics[width=\hsize]{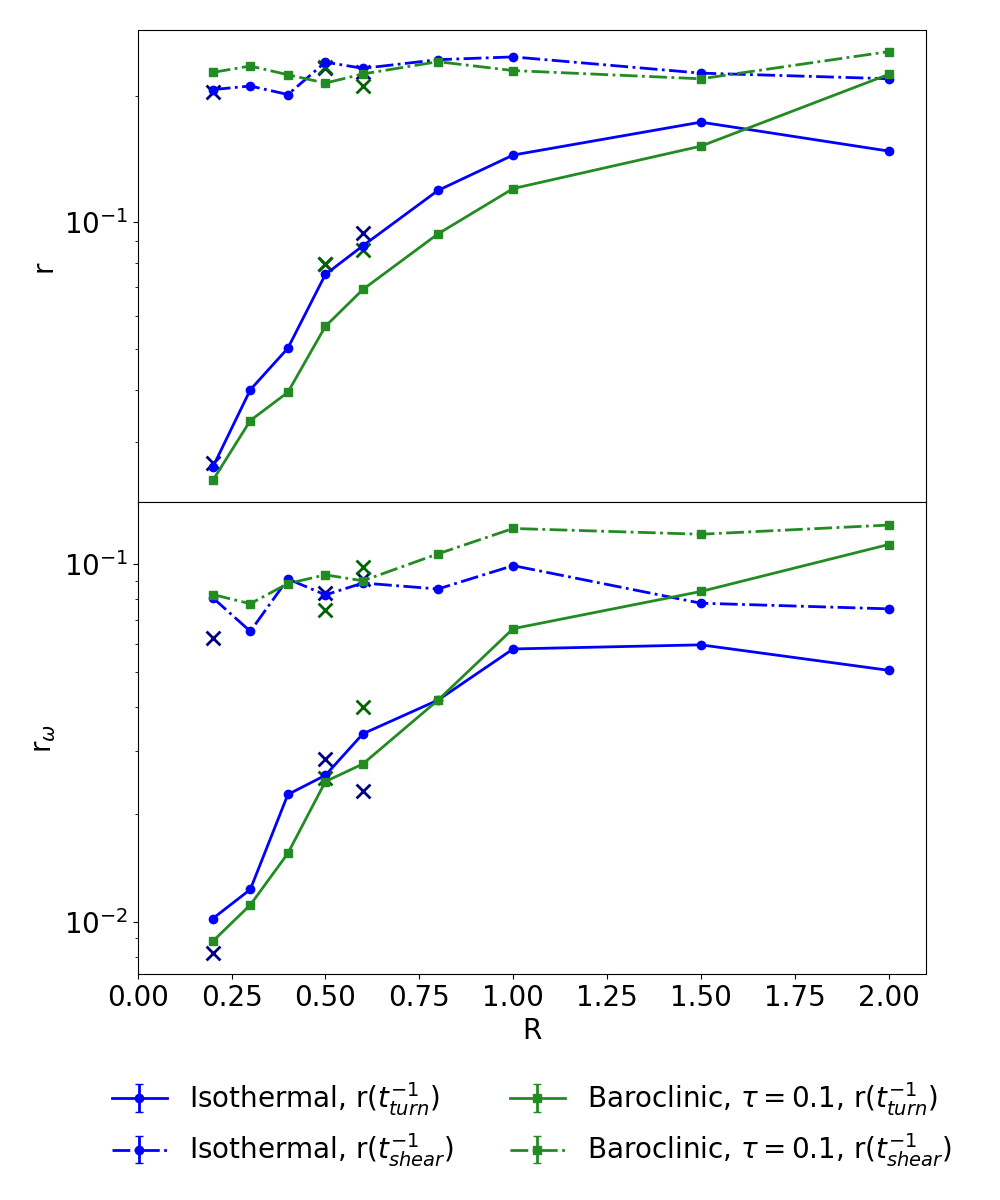}}
\caption{Vorticity and magnetic energy growth rates as a function of explosion width, $R$. The blue lines represent the isothermal models and the green ones the baroclinic cases. The dash-dotted lines correspond to growth rates in time units of t$_{shear}$ and solid lines in forcing turnover times, t$_{turn}$. Both time units are tabulated in Appendix \ref{App: Tabulated runs}. The points marked with crosses are the corresponding dynamo runs with $\mathbf{f_{\text{mom}}}$.}
\label{Fig: Growth rates vs R}
\end{figure}

In Fig. \ref{Fig: Timeseries growth rates} we plot the evolution of E$_{mag}$ and $\omega_{rms}$ for the isothermal runs with different $R$ and using the $f_{acc}$ forcing. The only model that does not develop a dynamo is that of the smallest forcing scale (i.e., $R=0.1$), and this is also true for the baroclinic case. Our interpretation is that, in this case, $Re$ is subcritical, and this allowed us to estimate a critical value of $Re \sim 40$ for the vorticity instability to take place. We find instead a critical magnetic Reynolds number of slightly less than 20 (see the tabulated values) for the dynamo instability.

In Fig. \ref{Fig: Growth rates vs R} we show the growth rates $r$ as a function of $R$ for both isothermal and baroclinic cases. We show the values of $r$ calculated using different time units: either $t_{shear}$ or $t_{turn}$. When choosing $t_{shear}$ as the time unit, $r$ approximately displays a constant behavior, while when choosing $t_{turn}$ the growth rates get vary considerably with $R$, as smaller expansion waves lead to much slower growths.

We observe that both magnetic and kinetic helicities grow in the dynamo cases, and start oscillating when $b_{rms}$ and $\omega_{rms}$ saturate. These oscillation resemble those seen in other instabilities such as the Tayler instability  \citep[e.g.,][]{Guerreroetal2019, Stefanietal2021, Monteiroetal2023}.

\subsection{Magnetic Prandtl number dependence}
\label{Section: Pm}
The magnetic Prandtl number (Pm) controls the scales at which the kinetic and magnetic energy cascades are truncated respectively by viscosity and resistivity. We varied Pm in our simulations so as to observe whether a difference in these truncation scales leads to a different behavior for the amplification of vorticity and magnetic field. In the ISM, Pm is usually much larger than 1 \citep{Ferriereetal2020}, which is different from the range explored in our models. We also explored a range of Pm slightly smaller than the unity, which is more typical of planetary environments.

We observe only a weak dependence on Pm for the models that do not develop a dynamo instability. In the isothermal case, we varied Pm from 0.25 up to 4 and see that \text{Re$_\omega$} either increases with Pm in the absence of rotation or slightly decrease with Pr when rotation is added (for more details see Table \ref{Table: Growth rates}). We recall here that \text{Re$_\omega$} is a dimensionless number introduced in \cite{EliasLopezetal2023} with the aim of quantifying the vorticity. In the baroclinic case, \text{Re$_\omega$} slightly decreases when Pm$=4$ compared to the case of Pm$=0.25$, independently on the presence of rotation. However, in the range of the explored values of Pm, we always report a decrease in the initial magnetic field.

Models with shear, conversely, develop a dynamo instability unless Pm$=4$ or above (see the tabulated values). We interpret this as a consequence of the lack of vorticity instability that does not develop when the physical viscosity increases above a certain value. The growth rates for the magnetic field increase with Pm, even in the cases where $\eta$ is kept constant and $\nu$ increased, of course up until the point where viscosity does not allow the hydro instability. Conversely, the growth of vorticity is approximately constant in the explored range.

\begin{table}[ht]
\centering
\caption{Vorticity and magnetic energy growth rates for different values of Pm.}
\begin{tabular}{@{}ccccccc@{}}
\hline \hline \\[-2.0ex]
Pm             & $\nu$ & $\eta$ & r ($t_{turn}^{-1}$) & r$_{\omega}$ ($t_{turn}^{-1}$)  \\ \hline
0.1            & 2$\cdot$10$^{-4}$    & 20$\cdot$10$^{-4}$    &  -                    & 10.52$\cdot$10$^{-3}$  \\
0.25           & 2$\cdot$10$^{-4}$    & 8$\cdot$10$^{-4}$     & 0.326$\cdot$10$^{-2}$ & 8.57$\cdot$10$^{-3}$   \\
0.5            & 2$\cdot$10$^{-4}$    & 4$\cdot$10$^{-4}$     & 1.22$\cdot$10$^{-2}$  & 9.33$\cdot$10$^{-3}$   \\
0.75           & 2$\cdot$10$^{-4}$    & 2.667$\cdot$10$^{-4}$ & 2.25$\cdot$10$^{-2}$  & 10.96$\cdot$10$^{-3}$  \\
1              & 2$\cdot$10$^{-4}$    & 2$\cdot$10$^{-4}$     & 2.78$\cdot$10$^{-2}$  & 10.22$\cdot$10$^{-3}$  \\
1.25           & 2.5$\cdot$10$^{-4}$  & 2$\cdot$10$^{-4}$     & 2.86$\cdot$10$^{-2}$  & 9.18$\cdot$10$^{-3}$   \\
1.5            & 3$\cdot$10$^{-4}$    & 2$\cdot$10$^{-4}$     & 3.07$\cdot$10$^{-2}$  & 10.01$\cdot$10$^{-3}$  \\
2              & 4$\cdot$10$^{-4}$    & 2$\cdot$10$^{-4}$     & 3.44$\cdot$10$^{-2}$  & 9.38$\cdot$10$^{-3}$   \\
4              & 8$\cdot$10$^{-4}$    & 2$\cdot$10$^{-4}$     &  -                    &  -                     \\ 
10             & 20$\cdot$10$^{-4}$   & 2$\cdot$10$^{-4}$     &  -                    &  -                     \\ \hline \hline
\end{tabular}
\label{Table: Growth rates}
\tablefoot{We show the values for $\nu$ and $\eta$, to better illustrate the different growths. We have not let the physical diffusivities reach numerical ones. The other diagnostics can be seen in Table \ref{Tab: Shear runs}.}
\end{table}

We find that for Pm$=0.1$ the vorticity is amplified, but, differently from the other cases, this instability is not followed by an exponential amplification of the magnetic field. This can be seen as a consequence of having a \text{Rm} below the critical value, since, in general, it is also possible to excite a dynamo at a Pm value of less than 0.1 \citep[e.g.,][]{Warneckeetal2023}. Our results are therefore consistent with what found in literature.

\subsection{Dependence on the cooling time}
\label{Section: cooling}

\begin{figure}[h]
\centerline{\includegraphics[width=\hsize]{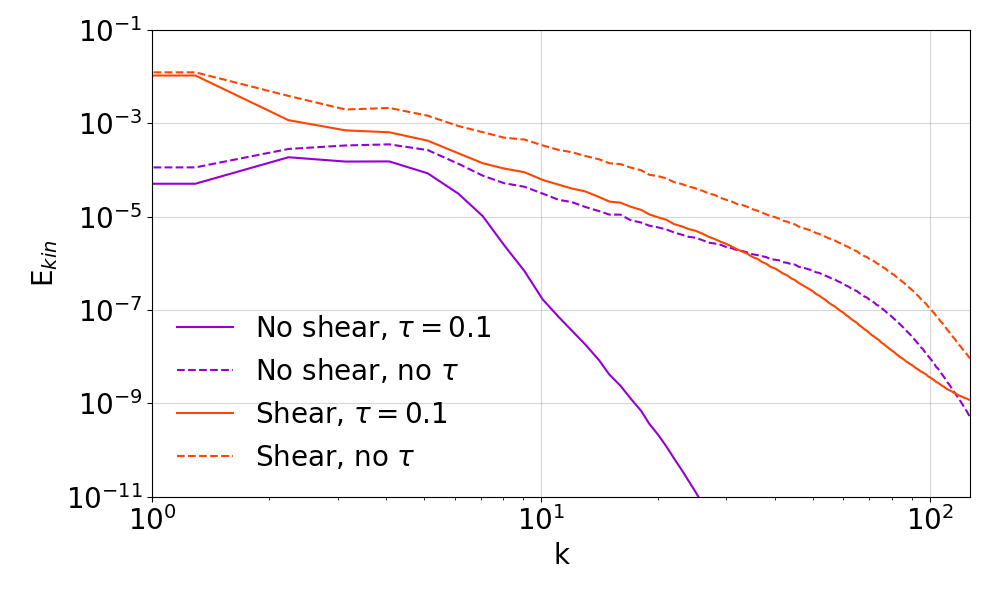}}
\caption{Kinetic spectra for models with $R=0.5$ and an optional shear and cooling term. The models with shear are shown after dynamo growth.}
\label{fig:spectra_tau}
\end{figure}

\noindent When the isothermal condition is relaxed, we let the temperature evolve according to Eq. \ref{Entropy_equation_Pencil}, where we use a Newtonian cooling term, regulated by the timescale $\tau_{cool}$. The use of the cooling function leads to spectra cut at short wavelengths in models that does not develop instabilities. This is shown in Fig. \ref{fig:spectra_tau} (compare the two purple lines). Differently, in the presence of shear and hence after a dynamo is excited, all the spectra recover their small-scale contribution and show again a wide dynamical range decreasing in a k$^{-2}$ fashion down to a dissipation scale, regardless of the cooling term (compare the orange curves). The use of the cooling term slightly diminishes the total amount of energy, but does not change notably the shape of the spectral distribution. However, we observe that in the presence of this cooling term the instability kicks in at much earlier times, independently of the value of $\tau_{cool}$, at least within the explored range. This can be seen as a quicker injection of vorticity in the system due to the cooling function. We also observe that the average angle between $\mathbf{\nabla}T$ and $\mathbf{\nabla}s$ slightly increases when a cooling function is used, hence leading to a larger contribution of the baroclinic term in seeding the vorticity.

Another possible interpretation is to invoke an effect similar to what observed by \cite{Raedleretal2011}. Irrotationally forced flows present peculiarities such as the possibility of having a negative magnetic diffusivity contribution from turbulent flows, especially at low Reynolds numbers, as shown analytically by \cite{Krause&Radler1980}, \cite{Radleretal2007} and \cite{Raedleretal2011}. The contributions to the diffusivity coming from the turbulence may affect the occurrence of dynamo instability even when one moves toward higher values of the magnetic Reynolds number, a regime that is closer to that of astrophysical bodies. \cite{Raedleretal2011} found, with mean-field approaches based on the second-order correlation approximation (SOCA), that a negative contribution to the magnetic diffusivity can come from the presence of turbulence in irrotational flows in the case of small Péclet numbers $Pe= u L / k$, where $u$, $L$, and $k$ are the typical velocity, length scale, and diffusivity of a system. In our case, the presence of a cooling time introduces an additional diffusion term for thermodynamical quantities, which results in $Rm$ being smaller  than it would be in the absence of cooling. However, the SOCA approximation is valid only for small magnetic Reynolds numbers, which is a condition that is not satisfied in our models.

Although our cooling function is not meant to model any specific astrophysical environment, we can attempt a comparison with typical values of the cooling in the ISM. Using the hydrogen cooling function $\Lambda$ \citep[e.g.,][]{Sutherland&Dopital1993} and assuming a temperature of 10$^4$ K,

\begin{equation*}
\begin{aligned}
    \log\frac{\Lambda}{\text{erg cm}^3 \text{s}^{-1}}(T=10^4 \text{K}) \approx -22 \quad \quad ; \quad \quad \frac{\partial e}{\partial t} = ... - \Lambda n \\
    \rightarrow \quad \tau_{cool} = \frac{k_B T}{\Lambda n}
    %= \frac{1.38\cdot10^{-6} \text{erg K}^{-1} 10^4 K}{10^{-22} \text{erg} \, \text{cm}^3 \text{s}^{-1} 1 \text{H cm}^{-3}} 
    \approx 10^{10}s \approx 0.5~{\rm kyr.}
\end{aligned}
\end{equation*}
If, as done by \cite{EliasLopezetal2023}, we considered a time unit of 8 Myr, then the cooling time $\tau_{cool}$ should be on the order of 0.001. In our models we can reach down to $\tau \sim 0.01$ for numerical stability reasons. This is one order of magnitude lower than typical ISM values (i.e., lower cooling rates and higher temperature differences than a one-phase ISM).

\subsection{Vorticity source terms}

\noindent From Eq. \ref{Eq: Vorticity source terms} we can evaluate which are the most relevant terms in the vorticity equation for the different runs. We find that in the isothermal nonrotating runs there are no positive terms comparable to the viscous ones. This fact along side the decrease in vorticity growth with resolution (see Fig. \ref{Omega rms vs resolution}), makes us think that in such cases only numerical diffusive sources are in action, in agreement with what we already observed in \cite{MeeBrandenburg2006} and \cite{EliasLopezetal2023}. 

\begin{figure}[t]
\centerline{\includegraphics[width=\hsize]{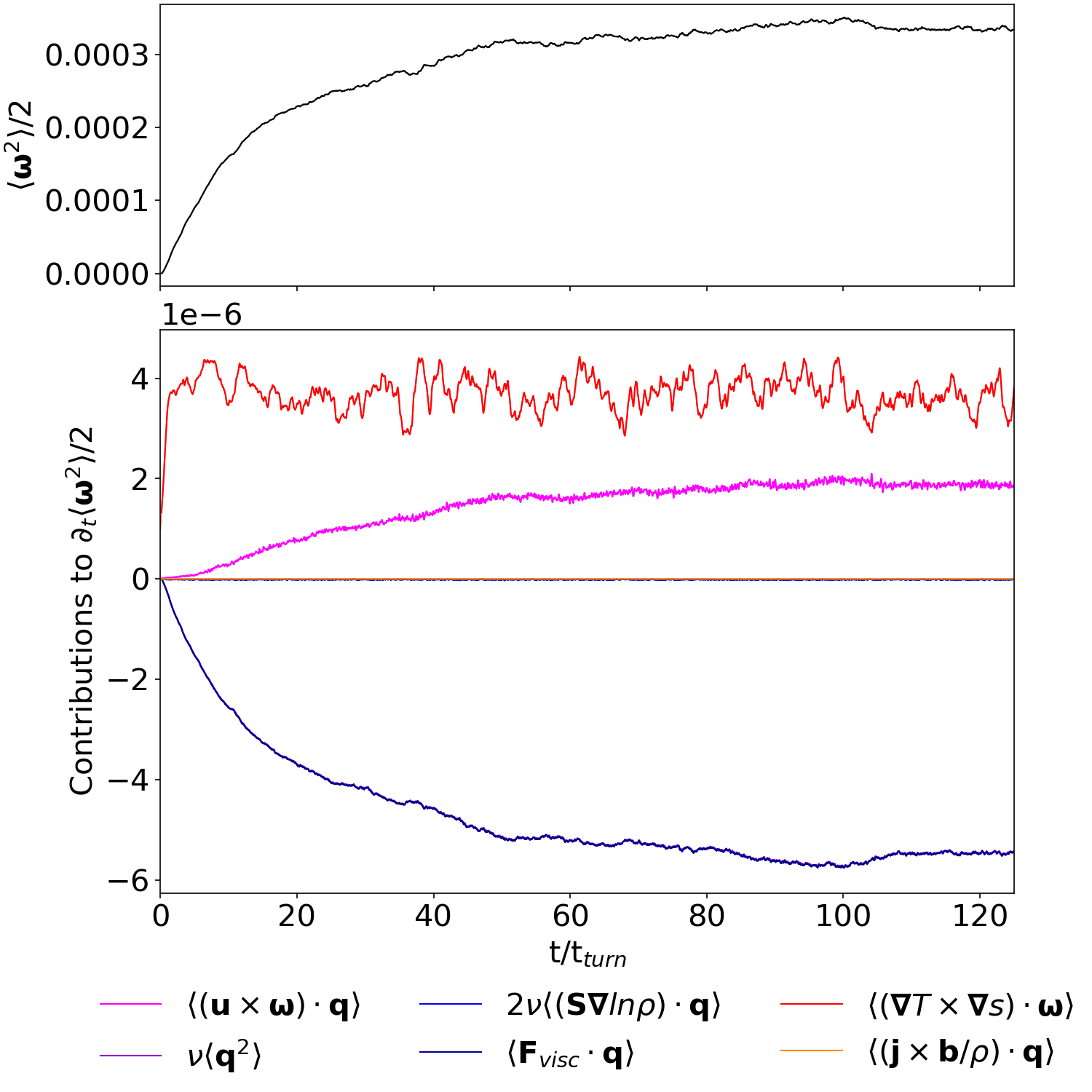}}
\caption{Time evolution for $\omega_{rms}^2/2$ (left axis and gray line) and vorticity growth terms (right axis and various colored lines for each term) for a non-isothermal run with R=0.5 and forcing in acceleration form (and by construction its contribution to $\langle \partial_t\boldsymbol{\omega}^2/2 \rangle$ is zero). This plot zooms in on the beginning of the temporal evolution of all the terms until $t\simeq$ 130 $t_{turn}$. However, the saturation regime does not present any relevant changes in time for more than 2000 $t_{turn}$, and no instability is reached. Both the $ 2\nu\langle \mathbf{S} \boldsymbol{\nabla} ln \rho \cdot \mathbf{q} \rangle$ and $\langle (\mathbf{j} \times \mathbf{b}/ \rho) \cdot \mathbf{q} \rangle$ terms are negligible and very close to 0, so this makes the total contribution of viscous forces, i.e., $\langle\mathbf{F}_{visc}\cdot\mathbf{q}\rangle$, overlap with $\nu \langle \mathbf{q}^2  \rangle$.}
\label{Fig: Vorticity source baroclinic}
\end{figure}

In Fig. \ref{Fig: Vorticity source baroclinic} we can see the time series of the different source terms as a function of time, for a representative non-isothermal case. All terms are very small. The baroclinic term dominates as the most positive contribution. As vorticity grows, the turbulent contribution $\langle (\mathbf{u}\times \boldsymbol{\omega}) \cdot \mathbf{q}\rangle$ gains importance, and the sum of both are counteracted by viscous forces, so that $\omega_{rms}$ saturates. From the viscous contributions, only $\nu \langle \mathbf{q}^2 \rangle$ is relevant, while $ 2\nu\langle \mathbf{S} \boldsymbol{\nabla} ln \rho \cdot \mathbf{q} \rangle$ is more than one order of magnitude lower. This last statement holds true for all the isothermal, non-isothermal and rotating, nonrotating cases. Obviously, the Lorentz term is irrelevant in the cases without dynamo, orders of magnitude below by comparison.

When the forcing is exactly irrotational, its corresponding term does not contribute to vorticity generation. But when it is applied in its second form (not exactly irrotational due to density fluctuations), there is indeed a small vorticity growth that leads to a similar behavior in strength and shape compared to the baroclinic source term. When applying this type of forcing in the non-isothermal runs, the forcing vorticity growth overtakes the baroclinic term to such an extent that this latter becomes negative.

\begin{figure}[t]
\centerline{\includegraphics[width=\hsize]{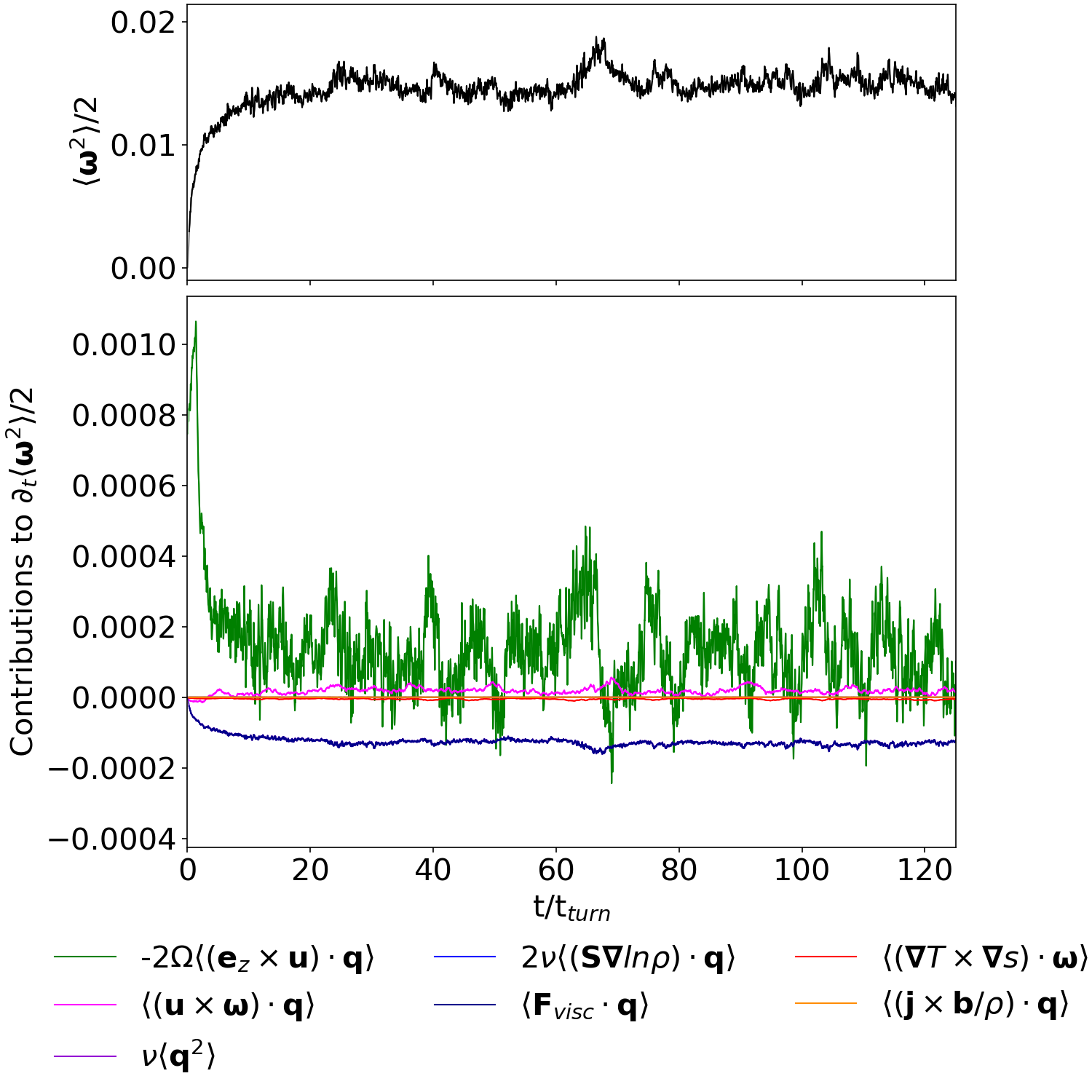}}
\caption{Time evolution for $\omega_{rms}^2/2$ and vorticity growth terms for a non-isothermal rotating run with R=0.5 and forcing in acceleration form. The notation is the same as in Fig. \ref{Fig: Vorticity source baroclinic}, and similarly $ 2\nu\langle \mathbf{S} \boldsymbol{\nabla} ln \rho \cdot \mathbf{q} \rangle$ and $\langle (\mathbf{j} \times \mathbf{b}/ \rho) \cdot \mathbf{q} \rangle$ overlap near 0, also making $\langle\mathbf{F}_{visc}\cdot\mathbf{q}\rangle$ overlap with $\nu \langle \mathbf{q}^2  \rangle$.}
\label{Fig: Vorticity source baroclinic rotation}
\end{figure}

When the rotation is included, the vorticity generation is more relevant. In this case, the Coriolis source creates an amount of vorticity that is later counteracted by viscous terms, so that the steady state is reached. In Fig. \ref{Fig: Vorticity source baroclinic rotation} we show the same plot as in Fig. \ref{Fig: Vorticity source baroclinic} but having added rotation. We can see that in the beginning the rotation term, $2\Omega\langle (\boldsymbol{e_z} \times \mathbf{u}) \cdot \mathbf{q} \rangle$, has a big positive spike that leads to the initial growth of vorticity. We note that it is larger than the baroclinic term in Fig. \ref{Fig: Vorticity source baroclinic} by more than one order of magnitude and oscillates substantially even becoming negative at times, leading to an overall a noisier $\omega_{rms}$.  
In the beginning of the run, the rotation contribution is mostly positive, and, when $\omega_{rms}$ saturates, the term is slightly positive in average, and of the same order of the viscous main contribution. Thus, in the presence of rotation, all other source terms become much less important.

\begin{figure}[t]
\centerline{\includegraphics[width=\hsize]{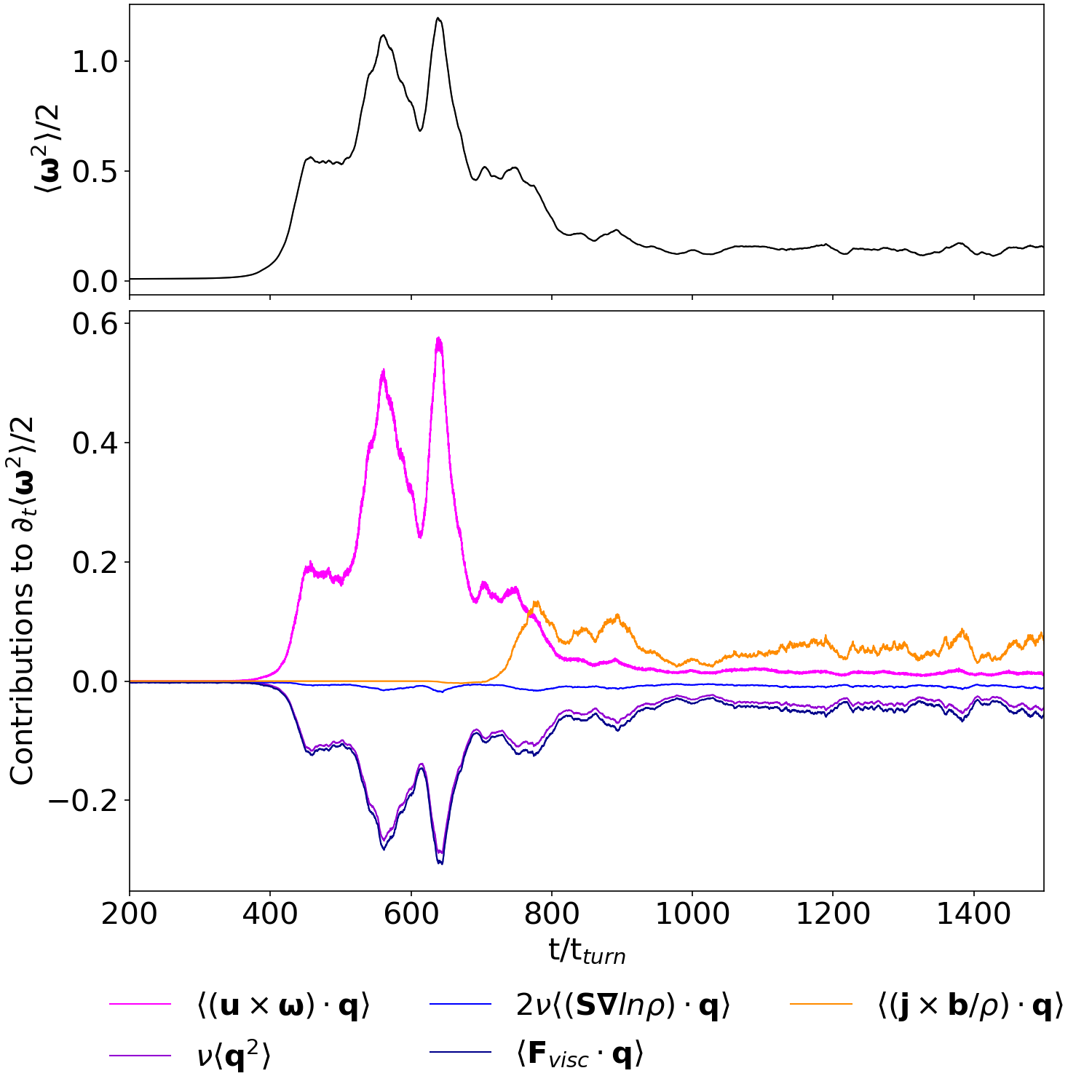}}
\caption{Time evolution for $\omega_{rms}^2/2$ and vorticity growth terms for a isothermal run with the presence of shear, hence leading to instability, and with $R=0.5$ and the exactly irrotational forcing. The notation is the same as in Fig. \ref{Fig: Vorticity source baroclinic}.}
\label{Fig: Vorticity source barotropic shear}
\end{figure}

For the cases with the shearing profile, both the HD and MHD instabilities make the contributions change substantially. In Fig. \ref{Fig: Vorticity source barotropic shear} we show all the relevant terms for an isothermal shearing run. As before, within the viscous forces, $\nu \langle \mathbf{q}^2\rangle$ still dominates over $2\nu\langle \mathbf{S} \boldsymbol{\nabla} ln \rho \cdot \mathbf{q} \rangle$, but the latter becomes more relevant in comparison to the cases described above.

The background shearing profiles increases $\omega_{rms}$ up to a certain values in a very few time steps. This value is kept approximatively until the vorticity instability kicks in (for example, until $t \simeq 400t_{turn}$ in Fig. \ref{Fig: Vorticity source barotropic shear}). Afterward, when vorticity is amplified, the advective term (which includes both shear and turbulence), $\langle (\mathbf{u}\times \boldsymbol{\omega}) \cdot \mathbf{q} \rangle$, brings the main positive contribution, as expected. The viscous forces are not enough to counteract this completely, thus leading to growth in $\omega_{rms}$.

The Lorentz term is negligible up to the point where dynamo starts. Then there is a brief time when it becomes slightly negative exactly when vorticity starts growing exponentially but still the dynamo has not kicked in. This behavior was also observed by \cite{Seta&Federrath2022}.
When the kinetic phase of the dynamo starts, the Lorentz term increases but as the Lorentz forces act against the flow, $\langle (\mathbf{u}\times \boldsymbol{\omega}) \cdot \mathbf{q} \rangle$ decreases more than $\langle (\mathbf{j} \times \mathbf{b}/ \rho) \cdot \mathbf{q} \rangle$ increases. This leads to a negative overall contribution and a decrease in vorticity, which later stabilizes at the end of the kinetic phase, and to the amplification of $B_y$ by winding. In all dynamo runs the Lorentz term always ends up surpassing the advective term $\langle (\mathbf{u}\times \boldsymbol{\omega}) \cdot \mathbf{q} \rangle$.

%The shearing term by itself contributes positively but in much less intensity than the other source terms. It also decreases when the magnetic instability kicks in. As with rotation, both the baroclinic and the forcing contributions remain small for the non-isothermal runs and the two forcing recipes. 

\section{Conclusion}

\noindent This work, a continuation of \cite{EliasLopezetal2023}, studied the vorticity and dynamo instability in the presence of irrotationally forced turbulence. We first explored the role played by the scale on which the irrotational forcing is acting. We find that, independent of the scale of this forcing, no dynamo instability develops for systems that do not include any shear. Also, while the root mean square values of vorticity weakly depend on the forcing scale, in no case do we observe an exponential amplification of vorticity. When shear is added to the picture, the vorticity is always exponentially amplified after a transient time if the kinematic viscosity is low enough, with the exception of the case of a very small forcing scale, which does not lead to any growth over thousands of turnover times. Then it is followed by a dynamo instability if the magnetic diffusivity is not too high. By analyzing kinematic spectra, we see how the typical scale of the system is provided by the forcing scale of the turbulence before the vorticity is amplified and, conversely, by the scale of the shear, in the saturation phase. Based on that, we observe that the growth rate of the dynamo depends on the scale of the expansion waves if the growth rate is calculated using the turnover time as the time unit, but it exhibits no $R$ dependence if the time unit is the shear timescale. The scale of the forcing also sets the time needed for the instability to develop. Models with a larger forcing scale amplify vorticity and magnetic fields after shorter times.  The only models that immediately develop the instability are those that include both the baroclinic term and a cooling function. We observe an increase of one order of magnitude for the growth rate with the magnetic Prandtl number, which we varied between 0.1 and 10. We extrapolated a critical value for the magnetic Reynolds number of slightly less than 20.

With these results in hand, we conclude that the presence of shear remains the basic ingredient for triggering a dynamo instability when subsonic turbulence is driven by spherical expansion waves. Future work will have to take into account turbulence forced on more than one scale at the same time as well as the role played by plane waves, before moving toward more complex models that include density stratification and also take shocks and supersonic flows into account.

This work both confirms previous results and extends them to a very wide, previously unexplored range of parameters and using two different kinds of spherically symmetric forcing: one that is curl-free and one that is not and takes density fluctuations into account. With the exploration of parameters, we definitely confirm that the shear is a key ingredient for a vorticity dynamo, while rigid rotation alone is not able to trigger it, regardless of the magnitude of the Reynolds number, the forcing, and the EoS.

\label{sec:conclusions}

\begin{acknowledgements}
This work has been carried out within the framework of the doctoral program in Physics of the Universitat Autònoma de Barcelona and it is partially supported by the program Unidad de Excelencia María de Maeztu CEX2020-001058-M. DV and AE are supported by the European Research Council (ERC) under the European Union’s Horizon 2020 research and innovation programme (ERC Starting Grant "IMAGINE" No. 948582, PI: DV). FDS acknowledges support from a Marie Curie Action of the European Union (Grant agreement 101030103). The authors acknowledge support from ``Mar\'ia de Maeztu'' award to the Institut de Ciències de l'Espai (CEX2020-001058-M). We are grateful to Axel Brandenburg, Matthias Rheinhardt, Eva Ntormousi, Federico Stasyszyn, Andrea
Pallottini, Amit Seta and Claudia Soriano-Guerrero for fruitful discussions. We also want to acknowledge the whole Pencil Code community for support. AE gratefully acknowledges Scuola Normale Superiore for hospitality during October 2023.
We acknowledge the use the SCAYLE supercomputer of the Spanish Supercomputing Network, via project RES/BSC Call AECT-2023-2-0034 (PI FDS).
\end{acknowledgements}

\bibliographystyle{aa}
\bibliography{biblio}

\begin{appendix}

\begin{landscape}
\section{Simulations with sinusoidal shear} \label{App: Tabulated runs}

The following tables contain all the simulations included in this work, as well as the most relevant runs in \cite{EliasLopezetal2023} (marked in \textcolor{violet}{violet}). We note that the runs from the past article do not have all diagnostics. For all models we used B$_0$=10$^{-6}$ as the seed field, $\Delta$t=0.02 as interval between two different explosions, $c_{s0}^2$=1 in the case of non-isothermal runs and an amplitude of the shearing profile $A=$ 0.2 in the shearing cases.

\begin{table}[ht]
\centering
\scriptsize
\caption{Barotropic (i.e., isothermal) runs without shear.}
\begin{tabular}{@{}ccccccccccccccccccc@{}}
\hline \hline \\[-2.0ex]
256$^3$         & $\nu$   & $\eta$    & $\Omega$ & $\phi_0$ & R   & t$_{tot}$  & t$_{turn}$   & k$_{\omega}$/k$_{f}$      & Re (Rm)   & Re$_{\omega}$    &  u$_{rms}$  &  u$_{rot}$/u$_{tot}$    & r ($t_{turn}^{-1}$) & r$_{\omega}$ ($t_{turn}^{-1}$)     \\ \hline \\[-2.0ex]
M\_0W0.10      & 2$\cdot$10$^{-4}$  & 2$\cdot$10$^{-4}$ & 0  & 2.6   & 0.10 & 282.77  & 0.77 & 0.04158  & 16.15 & 0.671 & (6.458 $\pm$ 0.042) $\cdot$10$^{-2}$ & 0.019 & - & - \\ 
M\_0W0.20      & 2$\cdot$10$^{-4}$  & 2$\cdot$10$^{-4}$ & 0  & 1     & 0.20 & 1266.21 & 1.42 & 0.01823  & 35.33 & 0.644 & (7.067 $\pm$ 0.047) $\cdot$10$^{-2}$ & 0.014 & - & -  \\  
M\_0W0.30      & 2$\cdot$10$^{-4}$  & 2$\cdot$10$^{-4}$ & 0  & 0.52  & 0.30 & 1304.95 & 2.16 & 0.01749  & 52.12 & 0.911 & (6.949 $\pm$ 0.072) $\cdot$10$^{-2}$ & 0.011 & - & -  \\  
M\_0W0.40      & 2$\cdot$10$^{-4}$  & 2$\cdot$10$^{-4}$ & 0  & 0.36  & 0.40 & 1293.14 & 2.78 & 0.02089  & 72.06 & 1.505 & (7.21  $\pm$ 0.12) $\cdot$10$^{-2}$  & 0.0085 & - & -  \\ 
M\_0W0.50      & 2$\cdot$10$^{-4}$  & 2$\cdot$10$^{-4}$ & 0  & 0.24  & 0.50 & 1313.37 & 3.56 & 0.01908  & 86.91 & 1.658 & (6.95  $\pm$ 0.14) $\cdot$10$^{-2}$  & 0.0072 & - & -  \\  
M\_0W0.60      & 2$\cdot$10$^{-4}$  & 2$\cdot$10$^{-4}$ & 0  & 0.175 & 0.60 & 1233.70 & 4.39 & 0.01845  & 102.5 & 1.890 & (6.83  $\pm$ 0.20) $\cdot$10$^{-2}$  & 0.0051 & - & -  \\  
M\_0W0.80      & 2$\cdot$10$^{-4}$  & 2$\cdot$10$^{-4}$ & 0  & 0.118 & 0.80 & 1329.22 & 5.74 & 0.02050  & 139.9 & 2.861 & (6.98  $\pm$ 0.29) $\cdot$10$^{-2}$  & 0.0044 & - & -  \\  
M\_0W1.00      & 2$\cdot$10$^{-4}$  & 2$\cdot$10$^{-4}$ & 0  & 0.087 & 1.00 & 1339.16 & 7.03 & 0.02186  & 178.5 & 3.901 & (7.14  $\pm$ 0.43) $\cdot$10$^{-2}$  & 0.0041 & - & -  \\  
M\_0W1.50      & 2$\cdot$10$^{-4}$  & 2$\cdot$10$^{-4}$ & 0  & 0.048 & 1.50 & 1433.08 & 11.2 & 0.01528  & 253.7 & 3.890 & (6.77  $\pm$ 0.73) $\cdot$10$^{-2}$  & 0.0031 & - & -  \\
M\_0W2.00      & 2$\cdot$10$^{-4}$  & 2$\cdot$10$^{-4}$ & 0  & 0.031 & 2.00 & 1489.96 & 16.7 & 0.01550  & 304.5 & 4.713 & (6.09  $\pm$ 0.81) $\cdot$10$^{-2}$  & 0.0022 & - & -  \\  \hline  \\[-2.0ex]
M\_2W0.10      & 2$\cdot$10$^{-4}$  & 2$\cdot$10$^{-4}$ & 2  & 2.0   & 0.10 & 1028.6  & 0.84 & 0.1666  & 14.82 & 2.469 & (5.927  $\pm$ 0.021) $\cdot$10$^{-2}$ & 0.11  & - & -  \\
M\_2W0.10      & 2$\cdot$10$^{-4}$  & 2$\cdot$10$^{-4}$ & 2  & 1.5   & 0.10 & 1114.23 & 1.03 & 0.1640  & 12.13 & 1.990 & (4.854  $\pm$ 0.020) $\cdot$10$^{-2}$ & 0.11  & - & -  \\
M\_2W0.20      & 2$\cdot$10$^{-4}$  & 2$\cdot$10$^{-4}$ & 2  & 1     & 0.20 & 1276.66 & 0.85 & 0.3822  & 59.07 & 22.57 & (11.81  $\pm$ 0.12) $\cdot$10$^{-2}$  & 0.29  & - & -  \\
M\_2W0.30      & 2$\cdot$10$^{-4}$  & 2$\cdot$10$^{-4}$ & 2  & 0.52  & 0.30 & 1237.98 & 1.08 & 0.6644  & 104.5 & 69.41 & (13.94  $\pm$ 0.29) $\cdot$10$^{-2}$  & 0.42  & - & -  \\
M\_2W0.40      & 2$\cdot$10$^{-4}$  & 2$\cdot$10$^{-4}$ & 2  & 0.36  & 0.40 & 1225.51 & 1.28 & 0.8895  & 157.0 & 139.6 & (15.70  $\pm$ 0.50) $\cdot$10$^{-2}$  & 0.50  & - & -  \\
M\_2W0.50      & 2$\cdot$10$^{-4}$  & 2$\cdot$10$^{-4}$ & 2  & 0.24  & 0.50 & 1250.13 & 1.65 & 1.030   & 190.1 & 195.7 & (15.21  $\pm$ 0.45) $\cdot$10$^{-2}$  & 0.54  & - & -  \\
M\_2W0.60      & 2$\cdot$10$^{-4}$  & 2$\cdot$10$^{-4}$ & 2  & 0.175 & 0.60 & 1178.19 & 1.93 & 1.105   & 233.5 & 257.7 & (15.57  $\pm$ 0.54) $\cdot$10$^{-2}$  & 0.52  & - & -  \\
M\_2W0.80      & 2$\cdot$10$^{-4}$  & 2$\cdot$10$^{-4}$ & 2  & 0.118 & 0.80 & 1252.24 & 2.36 & 1.247   & 339.3 & 422.9 & (16.96  $\pm$ 0.39) $\cdot$10$^{-2}$  & 0.57  & - & -  \\
M\_2W1.00      & 2$\cdot$10$^{-4}$  & 2$\cdot$10$^{-4}$ & 2  & 0.087 & 1.00 & 1261.90 & 3.03 & 1.424   & 413.3 & 588.4 & (16.53  $\pm$ 0.38) $\cdot$10$^{-2}$  & 0.57  & - & -  \\
M\_2W1.50      & 2$\cdot$10$^{-4}$  & 2$\cdot$10$^{-4}$ & 2  & 0.048 & 1.50 & 1285.2  & 4.52 & 1.983   & 625.0 & 1246  & (16.7   $\pm$ 1.2) $\cdot$10$^{-2}$   & 0.55  & - & -  \\
M\_2W2.00      & 2$\cdot$10$^{-4}$  & 2$\cdot$10$^{-4}$ & 2  & 0.031 & 2.00 & 1329.72 & 6.47 & 1.845   & 790.1 & 1450  & (15.8   $\pm$ 2.3) $\cdot$10$^{-2}$   & 0.44  & - & -  \\  \hline  \\[-2.0ex]
M\_0W0.20\_Pm0.25  & 2$\cdot$10$^{-4}$  & 8$\cdot$10$^{-4}$ & 0  & 1.0   & 0.20 & 1288.08 & 1.42 & 0.01828 & 35.1 (8.776)  & 0.6417  & (7.021 $\pm$ 0.044) $\cdot$10$^{-2}$ & 0.014 & - & -  \\
M\_0W0.20\_Pm4     & 8$\cdot$10$^{-4}$  & 2$\cdot$10$^{-4}$ & 0  & 1.0   & 0.20 & 1305.14 & 1.62 & 0.00620 & 7.729 (30.92) & 0.04792 & (6.184 $\pm$ 0.049) $\cdot$10$^{-2}$ & 0.013 & - & -  \\
M\_2W0.20\_Pm0.25  & 2$\cdot$10$^{-4}$  & 8$\cdot$10$^{-4}$ & 2  & 1.0   & 0,20 & 1241.23 & 0.85 & 0.3814  & 59.30 (14.80) & 22.39   & (11.840 $\pm$ 0.098) $\cdot$10$^{-2}$ & 0.29 & - & -  \\ 
M\_2W0.20\_Pm4     & 8$\cdot$10$^{-4}$  & 2$\cdot$10$^{-4}$ & 2  & 1.0   & 0.20 & 1295.54 & 1.30 & 0.3036  & 9.566 (38.27) & 29.04   & (7.653 $\pm$ 0.051) $\cdot$10$^{-2}$ & 0.23  & - & -  \\  \hline \\[-2.0ex]
512$^3$         & $\nu$    & $\eta$   &  $\Omega$ & $\phi_0$ & R   & t$_{tot}$   & t$_{turn}$      & k$_{\omega}$/k$_{f}$      & Re (Rm)   & Re$_{\omega}$   &      u$_{rms}$  &  u$_{rot}$/u$_{tot}$    & r ($t_{turn}^{-1}$) & r$_{\omega}$ ($t_{turn}^{-1}$)   \\ \hline
\violet{M\_0W0.20\_512}  & 2$\cdot$10$^{-4}$  & 2$\cdot$10$^{-4}$ & 0  & 1     & 0.20 & 47.34 & 1.47 & 0.00581 & 34.02 & 0.1975 & (6.803  $\pm$ 0.041) $\cdot$10$^{-2}$  &  0.012 & - & -  \\
M\_0W0.60\_512           & 2$\cdot$10$^{-4}$  & 2$\cdot$10$^{-4}$ & 0  & 0.175 & 0.60 & 598.85 & 4.531 & 0.01148 & 99.41 & 1.141 & (6.627  $\pm$ 0.19) $\cdot$10$^{-2}$ & - & - & -  \\
%M\_0W0.80\_512          & 2$\cdot$10$^{-4}$  & 2$\cdot$10$^{-4}$ & 0  & 0.087 & 0.60 &  &  &  &  &  &  & - & -  \\
M\_2W0.60\_512           & 2$\cdot$10$^{-4}$  & 2$\cdot$10$^{-4}$ & 2  & 0.175 & 0.60 & 339.57 & 2.13 & 1.174 & 211.0 & 247.8 & (14.07  $\pm$ 0.22) $\cdot$10$^{-2}$ & - & - & -  \\\hline \hline \\[-2.0ex]
256$^3$         & $\nu$    & $\eta$      & $\Omega$ & $\phi_0$ & $\Delta$t & R   & t$_{turn}$      & k$_{\omega}$/k$_{f}$      & Re (Rm)   & Re$_{\omega}$         &  u$_{rot}$/u$_{tot}$      & r ($t_{turn}^{-1}$) & r$_{\omega}$ ($t_{turn}^{-1}$)        \\ \hline \\[-2.0ex]
\violet{M\_0}                  & 2$\cdot$10$^{-4}$ & 2$\cdot$10$^{-4}$    & 0   & 1     & 0.02         & 0.2   &  1.42    & 0.01820  & 35.3  & 0.64    & 0.014  & - & - \\
\violet{M\_0s}                 & 2$\cdot$10$^{-4}$ & 2$\cdot$10$^{-4}$    & 0   & 1     & 1            & 0.2   &  4.32    & 0.00061  & 11.6  & 0.007   & - & - & -      \\
\violet{M\_0c}                 & 2$\cdot$10$^{-4}$ & 2$\cdot$10$^{-4}$    & 0   & 1     & $\delta t$   & 0.2   &  1.46    & 0.01609  & 34.3  & 0.55    & - & - & -      \\
\violet{M\_0low$\dagger$}      & 2$\cdot$10$^{-5}$ & 2$\cdot$10$^{-5}$    & 0   & 1     & 0.02         & 0.2   &  3.05    & 0.5913   & 163.8 & 97.03   & - & - & -      \\
\violet{M\_0low\_F2$\dagger$}  & 2$\cdot$10$^{-5}$ & 2$\cdot$10$^{-5}$    & 0   & 2     & 0.02         & 0.2   &  1.03    & 1.4231   & 486.9 & 708.34  & - & - & -      \\
\violet{M\_0lowc}              & 2$\cdot$10$^{-5}$ & 2$\cdot$10$^{-5}$    & 0   & 1     & $\delta t$   & 0.2   &  1.38    & 0.08582  & 362.6 & 31.12   & - & - & -      \\
\violet{M\_0highcF10}          & 2$\cdot$10$^{-2}$ & 2$\cdot$10$^{-2}$    & 0   & 10    & $\delta t$   & 0.2   &  0.61    & 0.03443  & 0.81  & 0.028   & - & - & -      \\ \hline \\[-2.0ex]
\violet{M\_2}                  & 2$\cdot$10$^{-4}$ & 2$\cdot$10$^{-4}$    & 2   & 1     & 0.02         & 0.2   &  0.84    & 0.3809   & 59.2  & 22.55   & 0.292  & - & - \\
\violet{M\_2c}                 & 2$\cdot$10$^{-4}$ & 2$\cdot$10$^{-4}$    & 2   & 1     & $\delta t$   & 0.2   &  0.85    & 0.38179  & 58.5  & 22.35   & - & - & -      \\
\violet{M\_2W0.5$\dagger$}     & 2$\cdot$10$^{-4}$ & 2$\cdot$10$^{-4}$    & 2   & 1     & $\delta t$   & 0.5   &  0.29    & 0.64222  & 171.7 & 110.24  & - & - & -      \\
\violet{M\_2low$\dagger$}      & 2$\cdot$10$^{-5}$ & 2$\cdot$10$^{-5}$    & 2   & 1     & 0.02         & 0.2   &  2.91    & 1.60296  & 171.8 & 276.5   & - & - & -  \\  \hline \hline \\[-2.5ex]
\end{tabular}
\label{Tab: Barotropic runs}
\end{table}
\end{landscape}

\newpage
\begin{landscape}
\begin{table}[ht]
\centering
\scriptsize
\caption{Baroclinic (i.e., non-isothermal) runs without shear.}
\begin{tabular}{@{}cccccccccccccccccccc@{}}
\hline \hline \\[-2.0ex]
256$^3$         & $\nu$  & $\chi$ & $\eta$   &  $\tau_{cool}$   & $\Omega$ & $\phi_0$ & R  & t$_{tot}$ & t$_{turn}$      & k$_{\omega}$/k$_{f}$      & Re (Rm)   & Re$_{\omega}$      &  u$_{rms}$     &  u$_{rot}$/u$_{tot}$     & r ($t_{turn}^{-1}$) & r$_{\omega}$ ($t_{turn}^{-1}$)     \\ \hline \\[-2.0ex]
MB\_0W0.20\_notau   & 2$\cdot$10$^{-4}$  & 2$\cdot$10$^{-4}$ & 2$\cdot$10$^{-4}$ & 0.1  & 0 & 2     & 0.20 & 1015.53 & 2.25 & 0.0304 & 22.23 & 0.675 & (4.445 $\pm$ 0.037) $\cdot$10$^{-2}$ & 0.015 & - & -  \\
MB\_0W0.20          & 2$\cdot$10$^{-4}$  & 2$\cdot$10$^{-4}$ & 2$\cdot$10$^{-4}$ &  -   & 0 & 1     & 0.20 & 905.98  & 2.69 & 0.1550 & 18.60 & 2.883 & (3.720 $\pm$ 0.038) $\cdot$10$^{-2}$ & 0.036 & - & -  \\
MB\_0W0.30\_notau   & 2$\cdot$10$^{-4}$  & 2$\cdot$10$^{-4}$ & 2$\cdot$10$^{-4}$ & 0.1  & 0 & 1.04  & 0.30 & 970.32  & 2.52 & 0.0101 & 44.69 & 0.451 & (5.959 $\pm$ 0.067) $\cdot$10$^{-2}$ & 0.011 & - & -  \\
MB\_0W0.30          & 2$\cdot$10$^{-4}$  & 2$\cdot$10$^{-4}$ & 2$\cdot$10$^{-4}$ &  -   & 0 & 0.52  & 0.30 & 1079.47 & 3.77 & 0.1748 & 29.84 & 5.215 & (3.979 $\pm$ 0.062) $\cdot$10$^{-2}$ & 0.040 & - & -  \\
MB\_0W0.40\_notau   & 2$\cdot$10$^{-4}$  & 2$\cdot$10$^{-4}$ & 2$\cdot$10$^{-4}$ & 0.1  & 0 & 0.72  & 0.40 & 964.04  & 3.23 & 0.0137 & 61.88 & 0.847 & (6.19  $\pm$ 0.10) $\cdot$10$^{-2}$ & 0.0076 & - & -  \\
MB\_0W0.40          & 2$\cdot$10$^{-4}$  & 2$\cdot$10$^{-4}$ & 2$\cdot$10$^{-4}$ &  -   & 0 & 0.36  & 0.40 & 1183.27 & 5.04 & 0.1744 & 39.77 & 6.930 & (3.98  $\pm$ 0.14) $\cdot$10$^{-2}$ & 0.037 & - & -  \\
MB\_0W0.50\_notau   & 2$\cdot$10$^{-4}$  & 2$\cdot$10$^{-4}$ & 2$\cdot$10$^{-4}$ & 0.1  & 0 & 0.48  & 0.50 & 980.79  & 4.19 & 0.0142 & 74.68 & 1.060 & (5.97  $\pm$ 0.14) $\cdot$10$^{-2}$ & 0.0072 & - & -  \\
MB\_0W0.50          & 2$\cdot$10$^{-4}$  & 2$\cdot$10$^{-4}$ & 2$\cdot$10$^{-4}$ &  -   & 0 & 0.24  & 0.50 & 1234.23 & 6.45 & 0.1655 & 48.50 & 8.018 & (3.88  $\pm$ 0.13) $\cdot$10$^{-2}$ & 0.030 & - & -  \\
MB\_0W0.60\_notau   & 2$\cdot$10$^{-4}$  & 2$\cdot$10$^{-4}$ & 2$\cdot$10$^{-4}$ & 0.1  & 0 & 0.35  & 0.60 & 998.6   & 5.14 & 0.0142 & 87.66 & 1.248 & (5.84  $\pm$ 0.16) $\cdot$10$^{-2}$ & 0.0065 & - & -  \\
MB\_0W0.60          & 2$\cdot$10$^{-4}$  & 2$\cdot$10$^{-4}$ & 2$\cdot$10$^{-4}$ &  -   & 0 & 0.175 & 0.60 & 1251.19 & 7.53 & 0.1633 & 59.99 & 9.777 & (4.00  $\pm$ 0.17) $\cdot$10$^{-2}$ & 0.024 & - & -  \\
MB\_0W0.80\_notau   & 2$\cdot$10$^{-4}$  & 2$\cdot$10$^{-4}$ & 2$\cdot$10$^{-4}$ & 0.1  & 0 & 0.236 & 0.80 & 1009.31 & 6.73 & 0.0183 & 191.1 & 2.174 & (5.95  $\pm$ 0.25) $\cdot$10$^{-2}$ & 0.0050 & - & -  \\
MB\_0W0.80          & 2$\cdot$10$^{-4}$  & 2$\cdot$10$^{-4}$ & 2$\cdot$10$^{-4}$ &  -   & 0 & 0.118 & 0.80 & 1252.50 & 8.69 & 0.1732 & 92.52 & 15.95 & (4.63  $\pm$ 0.31) $\cdot$10$^{-2}$ & 0.022 & - & -  \\
MB\_0W1.00\_notau   & 2$\cdot$10$^{-4}$  & 2$\cdot$10$^{-4}$ & 2$\cdot$10$^{-4}$ & 0.1  & 0 & 0.174 & 1.00 & 1020.70 & 8.32 & 0.0216 & 150.6 & 3.247 & (6.03  $\pm$ 0.33) $\cdot$10$^{-2}$ & 0.0040 & - & -  \\
MB\_0W1.00          & 2$\cdot$10$^{-4}$  & 2$\cdot$10$^{-4}$ & 2$\cdot$10$^{-4}$ &  -   & 0 & 0.087 & 1.00 & 1248.69 & 9.80 & 0.1828 & 128.6 & 23.31 & (5.14  $\pm$ 0.47) $\cdot$10$^{-2}$ & 0.022 & - & -  \\
MB\_0W1.50\_notau   & 2$\cdot$10$^{-4}$  & 2$\cdot$10$^{-4}$ & 2$\cdot$10$^{-4}$ & 0.1  & 0 & 0.096 & 1.50 & 1077.97 & 12.4 & 0.0168 & 229.5 & 3.852 & (6.12  $\pm$ 0.65) $\cdot$10$^{-2}$ & 0.0029 & - & -  \\
MB\_0W1.50          & 2$\cdot$10$^{-4}$  & 2$\cdot$10$^{-4}$ & 2$\cdot$10$^{-4}$ &  -   & 0 & 0.048 & 1.50 & 1248.71 & 13.4 & 0.2479 & 214.2 & 51.83 & (5.71  $\pm$ 0.87) $\cdot$10$^{-2}$ & 0.027 & - & -  \\ 
MB\_0W2.00\_notau   & 2$\cdot$10$^{-4}$  & 2$\cdot$10$^{-4}$ & 2$\cdot$10$^{-4}$ & 0.1  & 0 & 0.062 & 2.00 & 1164.33 & 18.8 & 0.0146 & 272.3 & 4.028 & (5.45  $\pm$ 0.85) $\cdot$10$^{-2}$ & 0.0016 & - & -  \\
MB\_0W2.00          & 2$\cdot$10$^{-4}$  & 2$\cdot$10$^{-4}$ & 2$\cdot$10$^{-4}$ &  -   & 0 & 0.031 & 2.00 & 1259.42 & 19.0 & 0.2886 & 273.4 & 75.96 & (5.5   $\pm$ 1.1)  $\cdot$10$^{-2}$  & 0.020 & - & -  \\  \hline \\[-2.0ex]
MB\_0W0.60\_tau001 & 2$\cdot$10$^{-4}$  & 2$\cdot$10$^{-4}$ & 2$\cdot$10$^{-4}$ & 0.01  & 0 & 0.175 & 0.60 & crashed & - & - & - & - & - & - & - & -  \\ 
MB\_0W0.60\_tau002 & 2$\cdot$10$^{-4}$  & 2$\cdot$10$^{-4}$ & 2$\cdot$10$^{-4}$ & 0.02  & 0 & 0.175 & 0.60 & crashed & - & - & - & - & - & - & - & -  \\ 
MB\_0W0.60\_tau005 & 2$\cdot$10$^{-4}$  & 2$\cdot$10$^{-4}$ & 2$\cdot$10$^{-4}$ & 0.05  & 0 & 0.175 & 0.60 & 1224.46 & 5.71 & 0.1240 & 78.95 & 9.772 & (5.26 $\pm$ 0.22) $\cdot$10$^{-2}$ & 0.016 & - & -  \\ 
MB\_0W0.60\_tau    & 2$\cdot$10$^{-4}$  & 2$\cdot$10$^{-4}$ & 2$\cdot$10$^{-4}$ & 0.1   & 0 & 0.175 & 0.60 & 1251.19 & 7.52 & 0.1633 & 59.99 & 9.777 & (3.99 $\pm$ 0.17) $\cdot$10$^{-2}$ & 0.024 & - & -  \\
MB\_0W0.60\_tau02  & 2$\cdot$10$^{-4}$  & 2$\cdot$10$^{-4}$ & 2$\cdot$10$^{-4}$ & 0.2   & 0 & 0.175 & 0.60 & 1264.19 & 9.56 & 0.1958 & 47.19 & 9.223 & (3.14 $\pm$ 0.13) $\cdot$10$^{-2}$ & 0.038 & - & -  \\ 
MB\_0W0.60\_tau05  & 2$\cdot$10$^{-4}$  & 2$\cdot$10$^{-4}$ & 2$\cdot$10$^{-4}$ & 0.5   & 0 & 0.175 & 0.60 & 1264.94 &10.38 & 0.1919 & 43.42 & 8.324 & (2.89 $\pm$ 0.10) $\cdot$10$^{-2}$ & 0.039 & - & -  \\
MB\_0W0.60\_tau1   & 2$\cdot$10$^{-4}$  & 2$\cdot$10$^{-4}$ & 2$\cdot$10$^{-4}$ & 1.0   & 0 & 0.175 & 0.60 & 1248.29 & 8.80 & 0.1520 & 51.21 & 7.774 & (3.41 $\pm$ 0.11) $\cdot$10$^{-2}$ & 0.026 & - & -  \\ \hline \\[-2.0ex]
MB\_2W0.20         & 2$\cdot$10$^{-4}$  & 2$\cdot$10$^{-4}$ & 2$\cdot$10$^{-4}$ & 0.1   & 2 & 2     & 0.20 & 891.15  & 2.10 & 0.4497 & 23.82 & 10.71 & (4.764 $\pm$ 0.035) $\cdot$10$^{-2}$  & 0.42 & - & -  \\
MB\_2W0.30         & 2$\cdot$10$^{-4}$  & 2$\cdot$10$^{-4}$ & 2$\cdot$10$^{-4}$ & 0.1   & 2 & 1.04  & 0.30 & 1062.24 & 2.62 & 0.5963 & 42.92 & 25.59 & (5.722 $\pm$ 0.060) $\cdot$10$^{-2}$  & 0.54 & - & -  \\
MB\_2W0.40         & 2$\cdot$10$^{-4}$  & 2$\cdot$10$^{-4}$ & 2$\cdot$10$^{-4}$ & 0.1   & 2 & 0.72  & 0.40 & 1108.78 & 2.82 & 0.6709 & 70.95 & 47.59 & (7.095 $\pm$ 0.11)  $\cdot$10$^{-2}$  & 0.60 & - & -  \\
MB\_2W0.50         & 2$\cdot$10$^{-4}$  & 2$\cdot$10$^{-4}$ & 2$\cdot$10$^{-4}$ & 0.1   & 2 & 0.48  & 0.50 & 1176.45 & 3.95 & 0.6898 & 79.28 & 54.66 & (6.342 $\pm$ 0.26)  $\cdot$10$^{-2}$  & 0.60 & - & -  \\
MB\_2W0.60         & 2$\cdot$10$^{-4}$  & 2$\cdot$10$^{-4}$ & 2$\cdot$10$^{-4}$ & 0.1   & 2 & 0.35  & 0.60 & 1190.03 & 4.70 & 0.6984 & 95.95 & 66.99 & (6.397 $\pm$ 0.20)  $\cdot$10$^{-2}$  & 0.62 & - & -  \\
MB\_2W0.80         & 2$\cdot$10$^{-4}$  & 2$\cdot$10$^{-4}$ & 2$\cdot$10$^{-4}$ & 0.1   & 2 & 0.236 & 0.80 & 1174.07 & 5.33 & 0.7139 & 150.8 & 107.5 & (7.541 $\pm$ 0.49)  $\cdot$10$^{-2}$  & 0.57 & - & -  \\
MB\_2W1.00         & 2$\cdot$10$^{-4}$  & 2$\cdot$10$^{-4}$ & 2$\cdot$10$^{-4}$ & 0.1   & 2 & 0.174 & 1.00 & 1160.27 & 5.65 & 0.7419 & 222.8 & 165.0 & (8.912 $\pm$ 0.77)  $\cdot$10$^{-2}$  & 0.60 & - & -  \\
MB\_2W1.50         & 2$\cdot$10$^{-4}$  & 2$\cdot$10$^{-4}$ & 2$\cdot$10$^{-4}$ & 0.1   & 2 & 0.096 & 1.50 & 1175.55 & 8.33 & 0.8140 & 343.3 & 278.5 & (9.154 $\pm$ 1.2)   $\cdot$10$^{-2}$  & 0.54 & - & -  \\ 
MB\_2W2.00         & 2$\cdot$10$^{-4}$  & 2$\cdot$10$^{-4}$ & 2$\cdot$10$^{-4}$ & 0.1   & 2 & 0.062 & 2.00 & 1194.15 & 12.6 & 0.8870 & 407.1 & 358.3 & (8.142 $\pm$ 1.2)   $\cdot$10$^{-2}$  & 0.43 & - & -  \\ \hline
MB\_0W\_Pm0.25     & 2$\cdot$10$^{-4}$  & 8$\cdot$10$^{-4}$ & 2$\cdot$10$^{-4}$ & 0.01  & 0 & 2.00  & 0.20 & 1129.75 & 1.29 & 0.0762 & 38.78 (9.694) & 2.954 & (7.755 $\pm$ 0.074) $\cdot$10$^{-2}$ & 0.019 & - & -  \\ 
MB\_0W\_Pm4        & 8$\cdot$10$^{-4}$  & 2$\cdot$10$^{-4}$ & 2$\cdot$10$^{-4}$ & 0.01  & 0 & 2.00  & 0.20 & 1154.10 & 2.77 & 0.0770 & 4.499 (18.00) & 0.346 & (3.600 $\pm$ 0.035) $\cdot$10$^{-2}$ & 0.022 & - & -  \\
MB\_2W\_Pm0.25     & 2$\cdot$10$^{-4}$  & 8$\cdot$10$^{-4}$ & 2$\cdot$10$^{-4}$ & 0.01  & 2 & 2.00  & 0.20 & 1138.86 & 2.11 & 0.450  & 23.65 (5.914) & 10.65 & (4.731 $\pm$ 0.033) $\cdot$10$^{-2}$ & 0.42  & - & -  \\ 
MB\_2W\_Pm4        & 8$\cdot$10$^{-4}$  & 2$\cdot$10$^{-4}$ & 2$\cdot$10$^{-4}$ & 0.01  & 2 & 2.00  & 0.20 & 1158.22 & 2.16 & 0.419  & 11.59 (23.17) & 4.858 & (4.635 $\pm$ 0.035) $\cdot$10$^{-2}$ & 0.39  & - & -  \\ \hline \\[-2.0ex]
512$^3$         & $\nu$  & $\chi$ & $\eta$   &  $\tau_{cool}$    &   $\Omega$ & $\phi_0$ & R & t$_{tot}$  & t$_{turn}$      & k$_{\omega}$/k$_{f}$      & Re (Rm)   & Re$_{\omega}$        &  u$_{rms}$        &  u$_{rot}$/u$_{tot}$     & r ($t_{turn}^{-1}$) & r$_{\omega}$ ($t_{turn}^{-1}$)   \\ \hline \\[-2.0ex]
\violet{MB\_0W0.20\_512}  & 2$\cdot$10$^{-4}$  & 2$\cdot$10$^{-4}$ & 2$\cdot$10$^{-4}$ & -    & 0 & 1     & 0.2  & 44.10   &  1.69  & 0.00717 & 29.6 & 0.21  & (5.884 $\pm$ 0.072) $\cdot$10$^{-2}$ & - & - & -  \\ 
MB\_0W0.60\_512           & 2$\cdot$10$^{-4}$  & 2$\cdot$10$^{-4}$ & 2$\cdot$10$^{-4}$ & 0.1  & 0 & 0.35  & 0.60 & 199.97  &  7.10  & 0.09355 & 63.5 & 5.92  & (4.23  $\pm$ 0.20) $\cdot$10$^{-2}$ & - & - & -  \\ 
MB\_2W0.60\_512           & 2$\cdot$10$^{-4}$  & 2$\cdot$10$^{-4}$ & 2$\cdot$10$^{-4}$ & 0.1  & 2 & 0.35  & 0.60 & 199.18  &  4.69  & 0.7196  & 96.1 & 69.1  & (6.40  $\pm$ 0.26) $\cdot$10$^{-2}$ & - & - & -  \\\hline \hline \\[-2.0ex]
256$^3$        & $\nu$    & $\chi$   & $\eta$  &  $\tau_{cool}$  & $\Omega$ & $\phi_0$ & $\Delta$t & R  & t$_{tot}$  & t$_{turn}$      & k$_{\omega}$/k$_{f}$      & Re (Rm)  & Re$_{\omega}$   &  u$_{rot}$/u$_{tot}$     & r ($t_{turn}^{-1}$) & r$_{\omega}$ ($t_{turn}^{-1}$)     \\ \hline \\[-2.0ex]
\violet{MB\_0}             & 2$\cdot$10$^{-4}$    & 2$\cdot$10$^{-4}$ & 2$\cdot$10$^{-4}$  & -     & 0   & 1   & 0.02        & 0.2    & - & 1.65  & 0.0878  &  30.3  & 0.27     & - & - & - \\
\violet{MB\_0c}            & 2$\cdot$10$^{-4}$    & 2$\cdot$10$^{-4}$ & 2$\cdot$10$^{-4}$  & -     & 0   & 1   & $\delta t$  & 0.2    & - & 1.65  & 0.00876 &  30.3  & 0.27     & - & - & - \\
\violet{MB\_0highcF10}     & 2$\cdot$10$^{-2}$    & 2$\cdot$10$^{-4}$ & 2$\cdot$10$^{-2}$  & -     & 0   & 10  & $\delta t$  & 0.2    & - & 0.64  & 0.01376 &  0.78  & 0.0108   & - & - & - \\
\violet{MB\_0lowc}         & 5$\cdot$10$^{-5}$    & 2$\cdot$10$^{-4}$ & 5$\cdot$10$^{-5}$  & -     & 0   & 1   & $\delta t$  & 0.2    & - & 1.61  & 0.02876 & 124.1  & 3.57     & - & - & - \\ 
\violet{MB\_0low}          & 5$\cdot$10$^{-5}$    & 2$\cdot$10$^{-4}$ & 5$\cdot$10$^{-5}$  & -     & 0   & 1   & 0.02        & 0.2    & - & 1.61  & 0.03046 & 124.4  & 3.79     & - & - & - \\
\violet{MB\_0low2c}        & 2$\cdot$10$^{-5}$    & 2$\cdot$10$^{-4}$ & 2$\cdot$10$^{-5}$  & -     & 0   & 1   & $\delta t$  & 0.2    & - & 1.60  & 0.0434  & 313.1  & 13.58    & - & - & - \\\hline \\[-2.0ex]
\violet{MB\_2}             & 2$\cdot$10$^{-4}$    & 2$\cdot$10$^{-4}$ & 2$\cdot$10$^{-4}$  & -     & 2   & 1   & 0.02        & 0.2    & - & 1.08  & 0.340   &  46.5  & 15.83    & - & - & - \\
\violet{MB\_2c}            & 2$\cdot$10$^{-4}$    & 2$\cdot$10$^{-4}$ & 2$\cdot$10$^{-4}$  & -     & 2   & 1   & $\delta t$  & 0.2    & - & 1.06  & 0.343   &  47.0  & 16.12    & - & - & - \\
\violet{MB\_2low$\dagger$} & 2$\cdot$10$^{-5}$    & 2$\cdot$10$^{-4}$ & 2$\cdot$10$^{-5}$  & -     & 2   & 1   & 0.02        & 0.2    & - & 0.96  &  0.6172 &  522.2 & 323.0    & - & - & - \\ \hline  \\[-2.0ex]
\violet{MB\_0W0.1}         & 2$\cdot$10$^{-2}$    & 2$\cdot$10$^{-2}$ & 2$\cdot$10$^{-2}$  & -     & 0   & 1   & 0.02        & 0.1    & - & 15.31 & 0.00076 & 0.0082 & 0.00001  & - & - & - \\
\violet{MB\_0W0.2}         & 2$\cdot$10$^{-2}$    & 2$\cdot$10$^{-2}$ & 2$\cdot$10$^{-2}$  & -     & 0   & 1   & 0.02        & 0.2    & - & 7.84  & 0.00419 & 0.0638 & 0.00027  & - & - & - \\
\violet{MB\_0W0.5}         & 2$\cdot$10$^{-2}$    & 2$\cdot$10$^{-2}$ & 2$\cdot$10$^{-2}$  & -     & 0   & 1   & 0.02        & 0.5    & - & 1.36  & 0.01923 & 0.9209 & 0.01772  & - & - & - \\
\violet{MB\_0W1}           & 2$\cdot$10$^{-2}$    & 2$\cdot$10$^{-2}$ & 2$\cdot$10$^{-2}$  & -     & 0   & 1   & 0.02        & 1.0    & - & 0.46  & 0.06042 & 5.399  & 0.3263   & - & - & - \\\hline \hline \\[-2.0ex]
\end{tabular}
\label{Tab: Baroclinic runs}
\end{table}
\end{landscape}

\newpage

\begin{landscape}

\begin{table}[ht]
\centering
\scriptsize
\caption{Models with sinusoidal shearing velocity profile. ND stands for the growth rates not determined, even though dynamo was present.}
\begin{tabular}{cccccccccccccccccccc}
\hline \hline \\[-2.0ex]
256$^3$            & $\nu$    & $\chi$ & $\eta$ &  $\tau_{cool}$  & $A$ & $\phi_0$ &  R & t$_{tot}$ & t$_{turn}$      & k$_{\omega}$/k$_{f}$      & Re(Rm)  & Re$_{\omega}$  & u$_{rot,0}$/u$_{tot}$   & u$_{rot}$/u$_{tot}$      & r ($t_{turn}^{-1}$) & r$_{\omega}$ ($t_{turn}^{-1}$)       \\ \hline \\[-2.0ex]
M\_S\_Pm0.1    & 2$\cdot$10$^{-4}$  & -  & 20$\cdot$10$^{-4}$  & -   & 0.20  & 1     & 0.2  & 4189.19 &  0.637  & 0.0915  & 78.5 (7.85) & 7.200 & 0.819 & 0.899 & - & 1.052$\cdot$10$^{-2}$ \\ 
M\_S\_Pm0.25   & 2$\cdot$10$^{-4}$  & -  & 8$\cdot$10$^{-4}$   & -   & 0.20  & 1     & 0.2  & 2277.66 &  0.635  & 0.0919  & 78.8 (19.7) & 7.296 & 0.811 & 0.894 & 3.261$\cdot$10$^{-3}$ & 8.573$\cdot$10$^{-3}$ \\ 
M\_S\_Pm0.5    & 2$\cdot$10$^{-4}$  & -  & 4$\cdot$10$^{-4}$   & -   & 0.20  & 1     & 0.2  & 3492.59 &  0.634  & 0.0912  & 78.9 (39.5) & 7.247 & 0.810 & 0.873 & 1.224$\cdot$10$^{-2}$ & 9.333$\cdot$10$^{-3}$ \\ 
M\_S\_Pm0.75   & 2$\cdot$10$^{-4}$  & -  &2.667$\cdot$10$^{-4}$& -   & 0.20  & 1     & 0.2  & 3927.51 &  0.634  & 0.0911  & 78.9 (59.2) & 7.221 & 0.809 & 0.867 & 2.254$\cdot$10$^{-2}$ & 1.096$\cdot$10$^{-2}$ \\ 
M\_S\_Pm1      & 2$\cdot$10$^{-4}$  & -  & 2$\cdot$10$^{-4}$   & -   & 0.20  & 1     & 0.2  & 3559.77 &  0.633  & 0.0910  & 78.9 (78.9) & 7.200 & 0.806 & 0.861 & 2.783$\cdot$10$^{-2}$ & 1.022$\cdot$10$^{-2}$ \\ 
M\_S\_Pm1.25   & 2.5$\cdot$10$^{-4}$& -  & 2$\cdot$10$^{-4}$   & -   & 0.20  & 1     & 0.2  & 3965.97 &  0.636  & 0.0910  & 62.9 (78.6) & 5.751 & 0.825 & 0.852 & 2.861$\cdot$10$^{-2}$ & 9.176$\cdot$10$^{-3}$ \\ 
M\_S\_Pm1.5    & 3$\cdot$10$^{-4}$  & -  & 2$\cdot$10$^{-4}$   & -   & 0.20  & 1     & 0.2  & 3811.68 &  0.637  & 0.0910  & 52.3 (78.5) & 4.767 & 0.817 & 0.865 & 3.073$\cdot$10$^{-2}$ & 1.001$\cdot$10$^{-2}$ \\ 
M\_S\_Pm2      & 4$\cdot$10$^{-4}$  & -  & 2$\cdot$10$^{-4}$   & -   & 0.20  & 1     & 0.2  & 4765.49 &  0.640  & 0.0914  & 39.0 (78.1) & 3.575 & 0.824 & 0.872 & 3.438$\cdot$10$^{-2}$ & 9.378$\cdot$10$^{-3}$ \\ 
M\_S\_Pm4      & 8$\cdot$10$^{-4}$  & -  & 2$\cdot$10$^{-4}$   & -   & 0.20  & 1     & 0.2  & 4757.68 &  0.649  & 0.0922  & 19.3 (77.0) & 1.775 & 0.848 & - & - & - \\ 
M\_S\_Pm10     & 20$\cdot$10$^{-4}$ & -  & 2$\cdot$10$^{-4}$   & -   & 0.20  & 1     & 0.2  & 3340.23 &  0.672  & 0.0947  & 7.44 (74.4) & 0.705 & 0.903 & - & - & - \\ \hline \\[-2.0ex]
M\_S\_W0.10    & 2$\cdot$10$^{-4}$  & -  & 2$\cdot$10$^{-4}$   & -   & 0.20  & 2     & 0.10 & 3012.18 &  0.33   & 0.04733 & 37.5   & 1.775 & 0.888 & - &  -  &  -  \\  
M\_S\_W0.20    & 2$\cdot$10$^{-4}$  & -  & 2$\cdot$10$^{-4}$   & -   & 0.20  & 1     & 0.20 & 3332.61 &  0.63   & 0.09071 & 78.8   & 7.103 & 0.809 & 0.855 &  2.616$\cdot$10$^{-2}$ & 1.022$\cdot$10$^{-2}$ \\ 
M\_S\_W0.30    & 2$\cdot$10$^{-4}$  & -  & 2$\cdot$10$^{-4}$   & -   & 0.20  & 0.52  & 0.30 & 4923.52 &  0.95   & 0.1372  & 118.2  & 15.98 & 0.818 & 0.861 &  3.997$\cdot$10$^{-2}$ & 1.233$\cdot$10$^{-2}$ \\   
M\_S\_W0.40    & 2$\cdot$10$^{-4}$  & -  & 2$\cdot$10$^{-4}$   & -   & 0.20  & 0.36  & 0.40 & 4867.96 &  1.26   & 0.1819  & 159.0  & 28.42 & 0.807 & 0.851 &  5.017$\cdot$10$^{-2}$ & 2.273$\cdot$10$^{-2}$ \\  
M\_S\_W0.50    & 2$\cdot$10$^{-4}$  & -  & 2$\cdot$10$^{-4}$   & -   & 0.20  & 0.24  & 0.50 & 4889.66 &  1.58   & 0.2275  & 197.5  & 44.37 & 0.818 & 0.861 &  7.514$\cdot$10$^{-2}$ & 2.572$\cdot$10$^{-2}$ \\  
M\_S\_W0.60    & 2$\cdot$10$^{-4}$  & -  & 2$\cdot$10$^{-4}$   & -   & 0.20  & 0.175 & 0.60 & 4092.11 &  1.90   & 0.2738  & 236.5  & 63.89 & 0.812 & 0.862 &  8.791$\cdot$10$^{-2}$ & 3.361$\cdot$10$^{-2}$ \\  
M\_S\_W0.80    & 2$\cdot$10$^{-4}$  & -  & 2$\cdot$10$^{-4}$   & -   & 0.20  & 0.118 & 0.80 & 4908.57 &  2.51   & 0.3607  & 319.3  & 113.5 & 0.812 & 0.846 &  1.190$\cdot$10$^{-1}$ & 4.175$\cdot$10$^{-2}$ \\  
M\_S\_W1.00    & 2$\cdot$10$^{-4}$  & -  & 2$\cdot$10$^{-4}$   & -   & 0.20  & 0.087 & 1.00 & 4918.58 &  3.08   & 0.4455  & 404.7  & 177.5 & 0.805 & 0.847 &  1.444$\cdot$10$^{-1}$ & 5.803$\cdot$10$^{-2}$ \\  
M\_S\_W1.50    & 2$\cdot$10$^{-4}$  & -  & 2$\cdot$10$^{-4}$   & -   & 0.20  & 0.048 & 1.50 & 1217.34 &  4.85   & 0.6884  & 579.9  & 399.1 & 0.804 & 0.840 &  1.729$\cdot$10$^{-1}$ & 5.959$\cdot$10$^{-2}$ \\ 
M\_S\_W2.00    & 2$\cdot$10$^{-4}$  & -  & 2$\cdot$10$^{-4}$   & -   & 0.20  & 0.031 & 2.00 & 1333.81 &  3.48   & 0.4909  & 359.4  & 176.4 & 0.942 & --    &  1.476$\cdot$10$^{-1}$ & 5.058$\cdot$10$^{-2}$ \\ \hline \\[-2.0ex]
%MB\_S\_W0.10   & 2$\cdot$10$^{-4}$ & 2$\cdot$10$^{-4}$  & 2$\cdot$10$^{-4}$    & 0.1 & 0.20  & 2     & 0.10 &  & & & & & & \\  
MB\_S\_W0.20   & 2$\cdot$10$^{-4}$ & 2$\cdot$10$^{-4}$  & 2$\cdot$10$^{-4}$    & 0.1 & 0.20  & 2     & 0.20 & 1710.04 & 0.68 & 01.102 & 73.00 & 8.046 & 0.97 & 0.97 & 2.433$\cdot$10$^{-2}$ & 8.832$\cdot$10$^{-3}$ \\ 
MB\_S\_W0.30   & 2$\cdot$10$^{-4}$ & 2$\cdot$10$^{-4}$  & 2$\cdot$10$^{-4}$    & 0.1 & 0.20  & 1.04  & 0.30 & 2307.15 & 1.02 & 01.628 & 110.2 & 17.82 & 0.96 & 0.96 & 3.377$\cdot$10$^{-2}$ & 1.116$\cdot$10$^{-2}$ \\   
MB\_S\_W0.40   & 2$\cdot$10$^{-4}$ & 2$\cdot$10$^{-4}$  & 2$\cdot$10$^{-4}$    & 0.1 & 0.20  & 0.72  & 0.40 & 2401.10 & 1.35 & 02.165 & 148.7 & 32.19 & 0.95 & 0.93 & 3.942$\cdot$10$^{-2}$ & 1.556$\cdot$10$^{-2}$ \\  
MB\_S\_W0.50   & 2$\cdot$10$^{-4}$ & 2$\cdot$10$^{-4}$  & 2$\cdot$10$^{-4}$    & 0.1 & 0.20  & 0.48  & 0.50 & 2444.66 & 1.68 & 02.590 & 185.9 & 48.15 & 0.95 & 0.92 & 5.656$\cdot$10$^{-2}$ & 2.470$\cdot$10$^{-2}$ \\  
MB\_S\_W0.60   & 2$\cdot$10$^{-4}$ & 2$\cdot$10$^{-4}$  & 2$\cdot$10$^{-4}$    & 0.1 & 0.20  & 0.35  & 0.60 & 2448.34 & 2.02 & 0.3010 & 223.3 & 67.21 & 0.96 & 0.93 & 6.921$\cdot$10$^{-2}$ & 2.768$\cdot$10$^{-2}$ \\  
MB\_S\_W0.80   & 2$\cdot$10$^{-4}$ & 2$\cdot$10$^{-4}$  & 2$\cdot$10$^{-4}$    & 0.1 & 0.20  & 0.236 & 0.80 & 2407.49 & 2.64 & 0.3889 & 302.9 & 117.8 & 0.94 & 0.92 & 9.382$\cdot$10$^{-2}$ & 4.174$\cdot$10$^{-2}$ \\  
MB\_S\_W1.00   & 2$\cdot$10$^{-4}$ & 2$\cdot$10$^{-4}$  & 2$\cdot$10$^{-4}$    & 0.1 & 0.20  & 0.174 & 1.00 & 2342.14 & 3.25 & 0.4744 & 384.4 & 182.3 & 0.93 & 0.91 & 1.202$\cdot$10$^{-1}$ & 6.614$\cdot$10$^{-2}$ \\  
MB\_S\_W1.50   & 2$\cdot$10$^{-4}$ & 2$\cdot$10$^{-4}$  & 2$\cdot$10$^{-4}$    & 0.1 & 0.20  & 0.096 & 1.50 & 2421.20 & 4.80 & 0.6864 & 586.3 & 402.2 & 0.92 & 0.88 & 1.516$\cdot$10$^{-1}$ & 8.403$\cdot$10$^{-2}$ \\ 
MB\_S\_W2.00   & 2$\cdot$10$^{-4}$ & 2$\cdot$10$^{-4}$  & 2$\cdot$10$^{-4}$    & 0.1 & 0.20  & 0.062 & 2.00 & 2421.66 & 6.44 & 0.9191 & 777.1 & 713.7 & 0.95 & 0.87 & 2.245$\cdot$10$^{-1}$ & 1.137$\cdot$10$^{-1}$ \\  \hline \\[-2.0ex]
MB\_S\_W0.50\_tau0005  & 2$\cdot$10$^{-4}$ & 2$\cdot$10$^{-4}$  & 2$\cdot$10$^{-4}$  & 0.005 & 0.20  & 0.48  & 0.50 & 1245.53 & 1.58 & 0.2232 & 198.4 & 44.28 & 0.81 & 0.77 & 5.863$\cdot$10$^{-2}$ & 3.137$\cdot$10$^{-2}$ \\
MB\_S\_W0.50\_tau001   & 2$\cdot$10$^{-4}$ & 2$\cdot$10$^{-4}$  & 2$\cdot$10$^{-4}$  & 0.01  & 0.20  & 0.48  & 0.50 & 1277.6  & 1.59 & 0.2258 & 196.1 & 44.27 & 0.84 & 0.80 & 5.993$\cdot$10$^{-2}$ & 2.776$\cdot$10$^{-2}$ \\
MB\_S\_W0.50\_tau05    & 2$\cdot$10$^{-4}$ & 2$\cdot$10$^{-4}$  & 2$\cdot$10$^{-4}$  & 0.05  & 0.20  & 0.48  & 0.50 & 1297.98 & 1.68 & 0.2362 & 187.5 & 44.28 & 0.92 & 0.88 & 7.864$\cdot$10$^{-2}$ & 3.164$\cdot$10$^{-2}$ \\
MB\_S\_W0.50\_tau01    & 2$\cdot$10$^{-4}$ & 2$\cdot$10$^{-4}$  & 2$\cdot$10$^{-4}$  & 0.1   & 0.20  & 0.48  & 0.50 & 1297.98 & 1.70 & 0.2410 & 183.7 & 44.28 & 0.95 & 0.93 & 6.769$\cdot$10$^{-2}$ & 3.251$\cdot$10$^{-2}$ \\
MB\_S\_W0.50\_notau    & 2$\cdot$10$^{-4}$ & 2$\cdot$10$^{-4}$  & 2$\cdot$10$^{-4}$  & -     & 0.20  & 0.48  & 0.50 & 979.59  & 1.57 & 0.2225 & 199.0 & 44.29 & 0.79 & 0.81 & 7.601$\cdot$10$^{-2}$ & 3.203$\cdot$10$^{-2}$ \\
MB\_S\_W0.20\_notau    & 2$\cdot$10$^{-4}$ & 2$\cdot$10$^{-4}$  & 2$\cdot$10$^{-4}$  & -     & 0.20  & 1     & 0.20 & 1769.62 & 0.65 & 0.0925 & 76.75 & 7.100 & 0.85 & 0.87 & 3.255$\cdot$10$^{-2}$ & 9.601$\cdot$10$^{-3}$ \\  \hline \hline \\[-2.0ex]
\end{tabular}
\label{Tab: Shear runs}
\end{table}

\end{landscape}

\newpage

\begin{landscape}

\begin{table}[ht]
\centering
\scriptsize
\caption{Continued.}
\begin{tabular}{cccccccccccccccccccc}
\hline \hline \\[-2.0ex]
512$^3$            & $\nu$    & $\chi$ & $\eta$ & $\tau_{cool}$  &  $A$ & $\phi_0$ & R  & t$_{tot}$ & t$_{turn}$      & k$_{\omega}$/k$_{f}$      & Re(Rm)  & Re$_{\omega}$  & u$_{rot,0}$/u$_{tot}$  & u$_{rot}$/u$_{tot}$     & r ($t_{turn}^{-1}$) & r$_{\omega}$ ($t_{turn}^{-1}$)     \\ \hline \\[-2.0ex]
\violet{M\_0A020\_512}   & 2$\cdot$10$^{-4}$ & -      & 2$\cdot$10$^{-4}$ &  -        & 0.20  & 1              & 0.2  & 1606.63 &  0.64  &   0.091 &  78.5  &  7.10  &  0.83  & - &  0.0264  & 0.0107  \\
M\_0A020\_512            & 1$\cdot$10$^{-4}$ & -      & 1$\cdot$10$^{-4}$ &  -        & 0.20  & 1              & 0.2  & 1950.99 &  0.64  &   0.090 &  157.0 &  14.2  &  - & - &  0.0348  &  0.0162  \\ 
M\_0A060\_512            & 2$\cdot$10$^{-4}$ & -      & 2$\cdot$10$^{-4}$ &  -        & 0.60  & 1              & 0.2  & 1642.89 &  1.98  &   0.281 &  227.3 &  6.39  &  - & - &  0.0995  &  0.0323 \\ \hline \hline \\[-2.0ex]
128$^3$            & $\nu$    & $\chi$ & $\eta$ &  $\tau_{cool}$   & $A$ & $\phi_0$  & R &  t$_{tot}$ & t$_{turn}$      & k$_{\omega}$/k$_{f}$      & Re(Rm)  & Re$_{\omega}$  & u$_{rot,0}$/u$_{tot}$   & u$_{rot}$/u$_{tot}$     & r ($t_{turn}^{-1}$) & r$_{\omega}$ ($t_{turn}^{-1}$)    \\ \hline \\[-2.0ex]
\violet{M\_0A000\_128}         & 2$\cdot$10$^{-4}$ & -    & 2$\cdot$10$^{-4}$  &  -  & 0     & 1        & 0.2  & - & - & - & 38.0  &  - & - & - & - &- \\ 
\violet{M\_0A002\_128}         & 2$\cdot$10$^{-4}$ & -    & 2$\cdot$10$^{-4}$  &  -  & 0.02  & 1        & 0.2  & - & - & - & 38.6  &  - & - & - & - &- \\  
\violet{M\_0A005\_128}         & 2$\cdot$10$^{-4}$ & -    & 2$\cdot$10$^{-4}$  &  -  & 0.05  & 1        & 0.2  & - & - & - & 41.9  &  - & - & - & - &- \\  
\violet{M\_0A010\_128}         & 2$\cdot$10$^{-4}$ & -    & 2$\cdot$10$^{-4}$  &  -  & 0.10  & 1        & 0.2  & - &  - & - & 52.1  &  - & - & -  & ND & ND \\
\violet{M\_0A020\_128}         & 2$\cdot$10$^{-4}$ & -    & 2$\cdot$10$^{-4}$  &  -  & 0.20  & 1        & 0.2  & - &  0.62   &  0.092 & 80.2  & 7.37   & 0.78  &  0.81 & 0.0296    & 0.00983 \\
\violet{M\_0A050\_128}         & 2$\cdot$10$^{-4}$ & -    & 2$\cdot$10$^{-4}$  &  -  & 0.50  & 1        & 0.2  & - &  - & - & 180.3 &  - & - & -  & ND & ND \\
\violet{M\_0A100\_128}         & 2$\cdot$10$^{-4}$ & -    & 2$\cdot$10$^{-4}$  &  -  & 1.00  & 1        & 0.2  & - &  - & - & 355.1 &  - & - & -  & ND & ND \\ \hline \\[-2.0ex]
\violet{M\_0A020\_W0.10\_128}  & 2$\cdot$10$^{-4}$ & -    & 2$\cdot$10$^{-4}$  &  -  & 0.20  & 1        & 0.1  & - & - & - & 36.1   & - & - & - & - & - \\ 
\violet{M\_0A020\_W0.15\_128}  & 2$\cdot$10$^{-4}$ & -    & 2$\cdot$10$^{-4}$  &  -  & 0.20  & 1        & 0.15 & - &  0.50   &  0.071 & 56.6  & 4.01   & 0.87  &  0.91  & 0.0286    & 0.00918 \\
\violet{M\_0A020\_W0.25\_128}  & 2$\cdot$10$^{-4}$ & -    & 2$\cdot$10$^{-4}$  &  -  & 0.20  & 1        & 0.25 & - &  0.74   &  0.112 & 104.9 & 11.76  & 0.73  &  0.76  & 0.0235    & 0.00853 \\ \hline \\[-2.0ex]
\violet{M\_0A020\_F0\_128}     & 2$\cdot$10$^{-4}$ & -    & 2$\cdot$10$^{-4}$  &  -  & 0.20  & 0        & -    & - & - & - & 70.5   &  - & - & - & - &- \\   
\violet{M\_0A020\_F0.5\_128}   & 2$\cdot$10$^{-4}$ & -    & 2$\cdot$10$^{-4}$  &  -  & 0.20  & 0.5      & 0.2  & - &  0.68   &  0.096 & 74.1  & 7.10   & 0.91  &  0.92  & 0.0290    & 0.00875 \\
\violet{M\_0A020\_F1.5\_128}   & 2$\cdot$10$^{-4}$ & -    & 2$\cdot$10$^{-4}$  &  -  & 0.20  & 1.5      & 0.2  & - &  0.58   &  0.099 & 85.8  & 8.49   & 0.70  &  0.75  & 0.0239    & 0.00857 \\ \hline \\[-2.0ex]
256$^3$            & $\nu$    & $\chi$ & $\eta$ &  $\tau_{cool}$  &   $A$ & $\phi_0$ & R  & t$_{tot}$ & t$_{turn}$      & k$_{\omega}$/k$_{f}$      & Re(Rm)  & Re$_{\omega}$  & u$_{rot,0}$/u$_{tot}$   & u$_{rot}$/u$_{tot}$     & r ($t_{turn}^{-1}$) & r$_{\omega}$ ($t_{turn}^{-1}$)     \\ \hline \\[-2.0ex]
\violet{M\_0A010}              & 2$\cdot$10$^{-4}$ & -      & 2$\cdot$10$^{-4}$    &  -   & 0.10      & 1       & 0.2 & -  &  1.00 &  0.073 & 49.8  &  3.61  & 0.51  &  0.68  & 0.0155    & 0.00756 \\
\violet{M\_0A015}              & 2$\cdot$10$^{-4}$ & -      & 2$\cdot$10$^{-4}$    &  -   & 0.15      & 1       & 0.2 & -  &  0.79 &  0.085 & 63.5  &  5.40  & 0.70  &  0.76  & 0.0220    & 0.00944 \\
\violet{H\_0A020}              & 2$\cdot$10$^{-4}$ & -      & -                    &  -   & 0.20      & 1       & 0.2 & -  &  - & - 78.9 & & - & - & - & - & - \\
\violet{M\_0A020}              & 2$\cdot$10$^{-4}$ & -      & 2$\cdot$10$^{-4}$    &  -   & 0.20      & 1       & 0.2 & -  &  0.63 &  0.091 & 78.9  &  7.19  & 0.81  &  0.86  & 0.0262    & 0.00969 \\
\violet{MB\_0A020}             & 2$\cdot$10$^{-4}$ & 2$\cdot$10$^{-4}$  & 2$\cdot$10$^{-4}$  &  - & 0.50  & 1   & 0.2 & -  &  0.65 &  0.093 & 76.9  &  7.15  & 0.85  &  0.88  & 0.0316    & 0.00925 \\
\violet{M\_0A050}              & 2$\cdot$10$^{-4}$ & -      & 2$\cdot$10$^{-4}$    &  -   & 0.50     & 1        & 0.2 & -  &  0.28 &  0.103 & 181.1 &  18.74 & 0.96  &  0.97  & 0.0510    & 0.0143  \\  \hline \hline
\end{tabular}
\label{Tab: Shear runs cont}
\end{table}

\end{landscape}

\end{appendix}

\end{document}